\documentclass[twocolumn,prd,superscriptaddress,nofootinbib,longbibliography,preprintnumbers]{revtex4-2}

\usepackage{graphicx}
\usepackage{amsmath,bm,amssymb,amsfonts,dsfont,mathtools}
\usepackage[usenames,dvipsnames]{xcolor}
\usepackage[normalem]{ulem}
\usepackage{url}
\usepackage{multirow}
\usepackage[colorlinks  = true,
            linkcolor   = NavyBlue,
            urlcolor    = NavyBlue,
            citecolor   = NavyBlue,
            anchorcolor = NavyBlue]{hyperref}

\usepackage{dcolumn}
\usepackage{bm}

\newcommand{\SF}{\texttt{skyFACT}}
\newcommand{\Opp}{\texttt{1pPDF}}

\newcommand{\dnds}{$dN/dS$}
\newcommand{\fluxunits}{ph cm$^{-2}$ s$^{-1}$ }
\newcommand{\paperI}{{\texttt{PaperI}}}

\newcommand{\SFnoGCE}{\texttt{sF-noGCE}}
\newcommand{\SFBBn}{\texttt{sF-B}}
\newcommand{\SFNFWn}{\texttt{sF-NFW126}}

\newcommand{\lsim}{\mathrel{\mathop{\kern 0pt \rlap
  {\raise.2ex\hbox{$<$}}}
  \lower.9ex\hbox{\kern-.190em $\sim$}}}
\newcommand{\gsim}{\mathrel{\mathop{\kern 0pt \rlap
  {\raise.2ex\hbox{$>$}}}
  \lower.9ex\hbox{\kern-.190em $\sim$}}}

\begin{document}

\preprint{LAPTH-008/24}

\title{The Galactic center excess at the highest energies:\\ morphology and photon-count statistics }%

\author{Silvia Manconi}
 \email{manconi@lapth.cnrs.fr}
 \affiliation{Laboratoire d'Annecy-le-Vieux de
Physique Théorique (LAPTh), CNRS, USMB, F-74940 Annecy, France}
\author{Francesca Calore}
\affiliation{Laboratoire d'Annecy-le-Vieux de
Physique Théorique (LAPTh), CNRS, USMB, F-74940 Annecy, France}
\author{Fiorenza Donato}

\affiliation{Department of Physics, University of Torino, via P. Giuria, 1, 10125 Torino, Italy}
\affiliation{Istituto Nazionale di Fisica Nucleare, via P. Giuria, 1, 10125 Torino, Italy}

\date{\today}

\begin{abstract}
The nature of the GeV gamma-ray Galactic center excess (GCE) in the data of Fermi-Large Area Telescope (LAT) is still to be unveiled. 
We present a new analysis of the inner Galaxy \textit{Fermi}--LAT data at energies above 10 GeV, 
based on an innovative method which combines the \SF~adaptive template fitting with and the \Opp~pixel-count statistics.
We find a strong evidence for the GCE also at high energies, $\sigma > 5$ regardless of the GCE spatial template.
Remarkably, our fits prefer the bulge morphological model over the dark matter  one at high significance, and show no evidence for an additional dark matter  template on top of the bulge component. 
Through the \Opp~analysis, we find that the model best describing the gamma-ray data 
 requires a smooth, diffuse GCE following a bulge morphology, together with sub-threshold point sources.
The \Opp~fit reconstructs a consistent population of faint point sources down at least to $10^{-12}$ \fluxunits. Between $10^{-12}$ \fluxunits and $10^{-11}$ \fluxunits the \Opp~measures a number of point sources significantly higher than the ones in the {\it Fermi} 4FGL catalog.
The robustness of our results brings further  support to the attempt of explaining, at least partially, the high-energy tail of the GCE in terms of a population of point sources, likely corresponding to millisecond pulsars. 

\end{abstract}

\maketitle

\section{Introduction}

The nature of the GeV gamma-ray Galactic center excess (GCE) in the data the he \textit{Fermi}-Large Area Telescope (LAT) \cite{Goodenough:2009gk,Vitale:2009hr,Hooper:2010mq,Hooper:2011ti,Abazajian:2012pn,Gordon:2013vta,Zhou:2014lva,Calore:2014xka,Daylan:2014rsa,TheFermi-LAT:2015kwa} is still to be unveiled. 
This unexpected component, detected significantly on top of the background coming from known sources, could be hinting at GeV thermal relic dark matter particles annihilating in the Galactic
dark matter halo \cite{Abazajian:2012pn,Daylan:2014rsa,Calore:2014nla,Agrawal:2014oha,Murgia:2020dzu,DiMauro:2021raz}. Albeit intriguing, the dark matter interpretation is currently challenged by a growing corpus of evidence that the morphology of the excess is better described by what expected from a population of millisecond pulsar (MSP)-like sources in the Galactic bulge \cite{Abazajian:2010zy,Bartels_bulgelum, Macias:2016nev, Macias:2019omb}. 
While the GCE photon flux is peaked at about few GeV, a high energy component extending up to tens of GeV has been reported by various studies. 
In particular, the high energy tail is found to be statistically significant no matter the nature of the spatial template used to describe the GCE emission, i.e. either when using a dark matter-like or stellar bulge-like excess template \cite{Calore:2014xka,Macias:2016nev,Bartels_bulgelum,DiMauro:2021raz}. 
The GCE high energy tail has been further characterized in e.g. \cite{Calore:2014nla,TheFermi-LAT:2015kwa,Fermi-LAT:2017opo}. The spectrum was found to be consistent among different the inner Galaxy (IG)~\cite{Calore:2014nla}, but the reconstructed properties resulted hampered by large systematic uncertainties, due to the intrinsic lower statistics and diffuse emission modeling.
Recently, a few template fitting analysis on updated \textit{Fermi}--LAT datasets found high-energy spectra consistent with earlier results~\cite{DiMauro:2021raz,Pohl:2022nnd,Cholis:2021rpp,McDermott2023}. 

If confirmed, such high-energy photons could be naturally explained by the inverse Compton scattering of electrons and positrons emitted by a population of MSPs in the Galactic bulge \cite{Petrovic:2014xra,Gautam:2021wqn}. 
For this putative population of unresolved MSPs, it would be crucial to know 
their phenomenological characteristics, such as the luminosity function, so to understand how many sources are expected to shine at different photon fluxes \cite{Dinsmore:2021nip}. Population studies often rely on observational-driven  quantities, such as the source-count distribution \dnds, i.e. the distribution of the number of sources $N$ as a function of their observed gamma-ray flux $S$, where $S$ is the detected flux integrated in a given energy range [$E_{\rm min}, E_{\rm max}$]. As for the \textit{Fermi}--LAT, the number of detected sources typically decreases for  $S \lsim 10^{-11}$~\fluxunits, where the detection efficiency drops below one 
\cite{Fermi-LAT:2015otn,DiMauro:2017ing}. 

The pioneering work of Ref.~\cite{2011ApJ...738..181M} indicated how to measure the \dnds~ below catalog flux threshold, and to decompose the different contributions to the gamma-ray sky by analyzing the statistics of gamma-ray photon counts in a pixelized sky map.   
To date, two implementations of the mathematical formalism introduced by Ref.~\cite{2011ApJ...738..181M} have been developed and applied to the \textit{Fermi}--LAT data: the so-called Non Poissonian Template Fitting (NPTF) \cite{Lee:2015fea,Mishra-Sharma:2016gis} and the \Opp~ \cite{Zechlin:1,Zechlin:2,Zechlin:3}.
The main applications of the method have led to measure the \dnds~ of isotropic, extragalactic sources \cite{Zechlin:1,DiMauro:2017ing,Lisanti:2016jub,Manconi:2019ynl}, and to contextually constrain dark matter annihilations at high latitudes \cite{Zechlin:3}. 

When these methods are applied to low Galactic latitudes, the modeling for the bright diffuse emissions could bias the results, as demonstrated in Refs.~\cite{Leane:2019xiy,Leane:2020nmi,Leane:2020pfc,Buschmann:2020adf}, casting doubts on early evidence for unresolved point sources to explain the GCE \cite{Lee:2015fea}. 
In \cite{Calore:2021jvg} (\paperI~in what follows) we pioneered a new approach combining \Opp~and the adaptive-constrained template fitting \SF~algorithm~\cite{Storm:2017arh, Bartels_bulgelum}.
More specifically, we used photon-count statistical methods within the \Opp~framework to measure faint point sources, well below the {\it Fermi}-LAT detection threshold in the IG, 
between 2 and 5 GeV. 
We found that all point sources and the 
bulge-correlated diffuse emission each contributes O(10\%)
of the total IG emission, thus disclosing a potential sub-threshold point-source contribution to the GCE at low energies. 
In addition, we showed that mis-modeling of the Galactic diffuse emission causing residual at low angular scales can lead to spurious evidence for new components, such as point sources, and we overcame this limitation by optimizing the background models through adaptive template fitting with \SF.
Complementary approaches have been recently explored, using spherical harmonics~\cite{Buschmann:2020adf} and Gaussian processes \cite{Mishra-Sharma:2021oxe}, as well as aiming at improving physics inputs relevant for the computation, 
e.g. the reconstructed gas maps~\cite{Pohl:2022nnd,Shmakov:2022vuc}. 
Moreover, new methodological developments have shown that 
machine learning techniques can be very powerful in discriminating a point-source vs.~diffuse origin of the GCE~\cite{List:2020mzd,2021PhRvD.104l3022L,2022MNRAS.516.2326B,Mishra-Sharma:2021oxe}.

At the highest energies, the nature of the GCE is even more challenging to probe. 
A recent dedicated analysis of the high-energy tail of the GCE ~\cite{Linden:2016rcf} 
analyzed 7 years of Pass8 \textit{Fermi}--LAT data with template fitting and NPTF. 
Using standard template fitting techniques, the authors find a statistically significant high-energy tail at $>10$~GeV with properties similar to the peak GCE emission at few GeV. An emission correlated with a GCE template was detected up to $\sim50$~GeV, even if above about $30$~GeV the morphology of the signal was found to be unconstrained. 
The NPTF analysis captured a mild evidence for a point-source explanation of the excess at energies above $\sim5$~GeV, although hampered by the same systematic uncertainties of the NPTF analysis in the 2--20~GeV energy bin \cite{Lee:2015fea}, as subsequently demonstrated in~\cite{Leane:2019xiy,Leane:2020nmi,Leane:2020pfc}. 

The goal of the current work is to assess the GCE significance 
and measure the characteristics of IG gamma rays at energies larger than 10~GeV.
We explore the source-count distribution of point-like sources and its spatial morphology. We will do so  by combining state-of-the-art methods as pioneered in \paperI. 
Our investigation of the GCE at high energies is novel in multiple aspects: 
(i) we adopt adaptive-constrained template fitting to probe the evidence of the
GCE and reconstruct its morphology at the highest energies; 
(ii) we combine this technique with pixel count statistical methods in order to assess the role of sub-threshold point sources to the GCE at $E>10$~GeV, while minimizing the mis-modelling of Galactic diffuse emission backgrounds; (iii) we characterize the properties of the gamma-ray emission in the IG at $E>10$~GeV by reconstructing the flux distribution of point sources  well below the Fermi-LAT detection threshold, and by estimating their spatial density as compared to other regions in the sky.

The paper is organized as follows. The methodology is introduced in section \ref{sec:methods}, followed by the definition of the \textit{Fermi}--LAT dataset and the region of interest used in section \ref{sec:data}.
The results of the \SF\ and \Opp\ analysis are illustrated in sections \ref{sec:SF_results} and \ref{sec:Opp_results}, before concluding in section \ref{sec:conclusions}. Additional systematic tests are summarized in the appendix.

\section{Methodology}\label{sec:methods}
The gamma-ray emission observed by \textit{Fermi}--LAT is usually separated in truly diffuse emissions and point-like (or extended) sources. 
Among diffuse emissions, the one of our own Galaxy, i.e. the Galactic diffuse emission, represents the dominant contribution to the observed photons, in particular towards the Galactic plane. It originates from hadronic and leptonic cosmic rays interacting with the interstellar gas and radiation fields. Other diffuse structures extended few tens of degrees in the sky are the \textit{Fermi} Bubbles~\cite{Fermi-LAT:2014sfa,Herold:2019pei} and the Loop~I~\cite{Wolleben:2007pq}.  

Individual sources of gamma rays, connected to e.g. compact Galactic objects such as pulsars, or distant active galaxies are collected in the source catalogs compiled by the LAT collaboration. The last catalog, the 4FGL-DR4 \cite{Fermi-LAT:2022byn,Ballet:2023qzs} contains 7194 sources significantly detected between energies of 50~MeV to 1~TeV with a significance larger than $5\sigma$ over a data-taking period of about 14 years. 
At high latitudes, sources that are too faint to be detected with a statistical significance of $5\sigma$ during a given survey, together with a residual cosmic-ray contamination, make up the diffuse gamma ray background~\cite{2015ApJ...799...86A} (DGRB), which exhibits anisotropies that could be used to characterize the different populations of dim sources contributing to it~\cite{Fermi-LAT:2018udj}.
These so-called unresolved sources can also considerably contribute to the gamma-ray sky in the inner part of the Galaxy. 

The methods, based on the \SF\ and the \Opp\ tools, both model diffuse and individual sources of gamma rays using complementary approaches. 
They are based on the use of templates, i.e. pixelized maps describing the energetic characteristics of different sky components as a function of the sky coordinates.
To this, the photon-count method adds a statistical description of bright and faint point sources. 
While the methods and procedures generally follow what introduced in  \paperI~ for the energy bin 2-5~GeV, we  
present here the developments that turned out to be necessary to extend our analysis to higher energies. 

\begin{figure*}[t]
\centering
\includegraphics[width=0.49\textwidth]{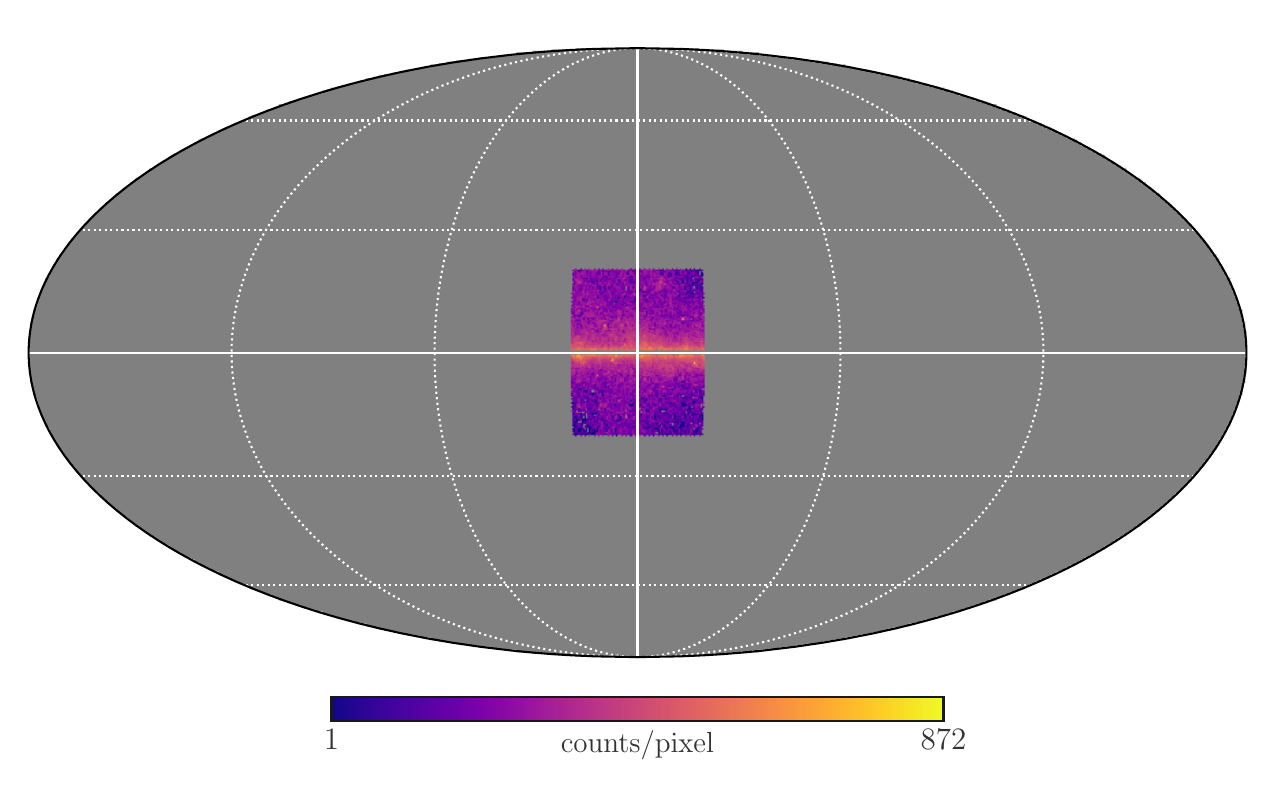}
\includegraphics[width=0.49\textwidth]{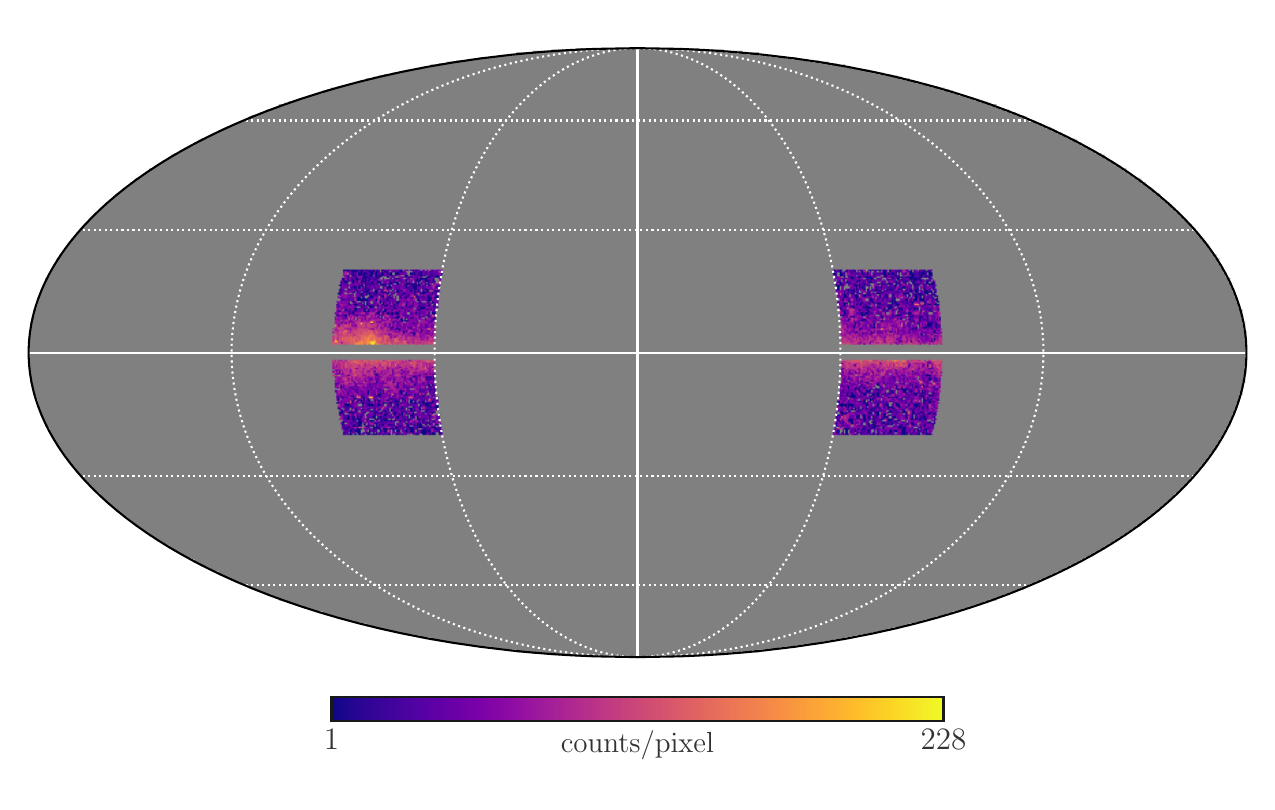}\\
\includegraphics[width=0.49\textwidth]{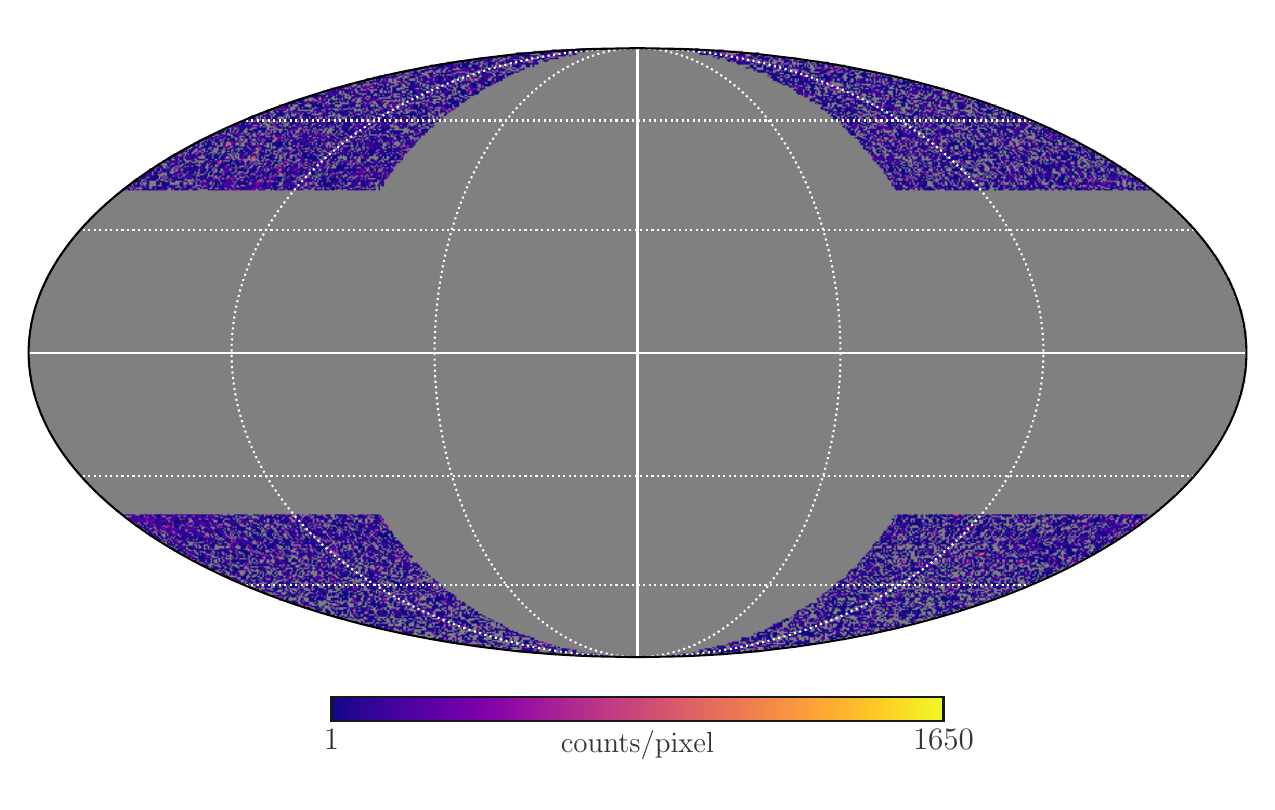}
\caption{\label{fig:counts} The \textit{Fermi}--LAT counts per pixel in the energy bin [10, 300]~GeV for \texttt{CLEAN} events as measured within the region of interest for the \Opp~analysis in the \textit{Inner Galaxy} (upper left panel),  \textit{Outer Galaxy} (upper right) and \textit{extragalactic sky} (lower panel). All maps are in Mollweide projection and  using HEALPix resolution parameter $\kappa=7$. }
\end{figure*}

\subsection{Adaptive-constrained template fitting with SkyFACT}
\label{sec:SF_intro}
The \SF\ method was introduced in Ref.~\cite{Storm:2017arh} and applied to the GCE in Ref.~\cite{Bartels_bulgelum}.  By introducing a large number of parameters, and using regularization condition for the likelihood, it accounts for intrinsic uncertainties in spectral and spatial predictions of the different templates entering the fit of \textit{Fermi}--LAT data. 

The modeling of the gamma-ray sky and statistical analysis closely follows Ref.~\cite{Bartels_bulgelum} and \paperI. 
The gamma-ray sky is interpreted as the combination of several model components, each of which is characterized by 
a spectral and a spatial (i.e.~morphology) input template, which are factorized in the \SF~implementation. 
The Galactic diffuse emission follows the implementation in Ref.~\cite{Bartels_bulgelum}: It is composed by 
an inverse Compton contribution whose spectrum and morphology are computed for a typical scenario of cosmic-ray sources and propagation parameters with the \texttt{DRAGON} code~\cite{2011ascl.soft06011M}, and 
a contribution from neutral pion decay split into three different templates for more flexibility. Each 
gas template corresponds, spatially, to a different gas ring (0 -- 3.5 kpc, 3.5 -- 6.5 kpc, 6.5 -- 19 kpc), and it is the sum of atomic and molecular hydrogen distribution within that ring, as available in the GALPROP public release\footnote{\url{https://galprop.stanford.edu/}} \cite{Porter:2021tlr}. The gamma-ray input spectrum of the gas component 
corresponds to the pion decay spectrum, it is taken from Ref.~\cite{2012ApJ...750....3A}, and it is the same for the three rings.   
On top of the Galactic diffuse emission, the model takes contribution from: An isotropic spatial component with the best-fit DGRB spectrum from Ref.~\cite{2015ApJ...799...86A} and a uniform spatial template; all point-like and extended 4FGL sources; the {\it Fermi}~bubbles with input spectrum from Ref.~\cite{Fermi-LAT:2014sfa} and a uniform geometrical template input morphology. Finally, we consider, on top of this \emph{background} model, a template for the GCE.

As for all other components, we must specify the input 
conditions of the spectral and spatial part of the GCE, separately.
In our setup, regardless of the spatial template, the baseline choice for the GCE input spectrum is a power law with index of -2.5. We will show that the choice of the spectral index does not affect the final results since we leave full freedom for spectral modulation, see below.
In \paperI~baseline run's configuration, the GCE input spectral model was instead an MSP-motivated spectrum, 
i.e.~a power-law with exponential cutoff~\cite{McCann:2014dea}.
However, this spectrum, being zero at high energies, cannot be modulated such to account for additional flux at $E>10$ GeV, so that it is not adapted for studying the presence of a high-energy tail of the GCE.
We discuss systematic uncertainties related to the GCE input spectrum in appendix~\ref{app:SF_sys}.
As in \paperI, for different choices of the input spectrum and independently on it, we test two different morphologies of the GCE: (i) a dark-matter inspired 
template following a generalised Navarro-Frenk-White dark matter density squared with inner slope 1.26 (NFW126); 
(ii) a bulge template, made up by a boxy-bulge component from Ref.~\cite{2013MNRAS.434..595C} and a nuclear bulge as in Ref.~\cite{Bartels_bulgelum}.
While a more recent determination of the bulge from Ref.~\cite{2020MNRAS.495.3350C} has been proposed to better fit the data~\cite{2024arXiv240102481Z, Song24prep}, 
we adopt the same bulge model as in \paperI, for the sake of comparing fairly with our previous results. We will nonetheless test also this new bulge model in appendix~\ref{app:SF_sys}.~%
We refer to Refs.~\cite{Storm:2017arh, Bartels_bulgelum,Calore:2021jvg} for more details about the gamma-ray model. 

\SF~implements adaptive-constrained template fitting by enabling each model component to vary both spectrally and spatially. Starting from the input spectra and morphologies of each model component, as mentioned above, \SF~minimizes a penalized likelihood composed by a standard Poissonian term and a regularization term 
which is governed by hyperparameters preventing overfitting and effectively constraining the modulation 
of the different components. 
Modulation parameters for the spectral and spatial parts of each component are therefore adjusted 
in order to maximize the likelihood. The result of the fit for each component is therefore a 
{\it modulated}, output spectrum and spatial template. How much this output differs from the input
templates depends on the freedom allowed during the fit. 
The minimization algorithm is the L-BFGS-B (Limited memory BFGS with Bound constraints) one. For more details about 
the technical implementation, we refer the reader to Ref.~\cite{Storm:2017arh}.
For this work, the baseline setup of hyperparameters follows \texttt{run5} of Ref.~\cite{Storm:2017arh} 
for all components but the additional GCE. Similarly to what done in \paperI, the GCE spectrum 
is allowed to fully vary in each energy bin independently, while its morphology is fixed to the 
input template, so that there is no additional freedom enabled for spatial modulation parameters.

The statistical framework implemented by \SF\ is purely frequentist. 
The output of the fit is the maximum likelihood value of the specific model considered. 
As such, we cannot properly do model comparison -- which would require a Bayesian approach -- within 
\SF. On the other hand, we can test the preference for additional model components
by considering a nested model, i.e.~a model which includes a new sky contribution (e.g.~GCE) on top
of another, less complex, model. The augmented model reduces to the other (i.e.~to the null hypothesis) for some
choice of the additional component model parameters.
For nested models the calculation of the evidence for any additional component is well defined, and 
follows the modified $\delta-\chi^2$ statistics as described in Ref.~\cite{Bartels_bulgelum}.
More specifically, we start by fitting the gamma-ray sky without a GCE diffuse template (\SFnoGCE). We proceed adding a GCE diffuse template following the bulge morphology (\SFBBn) or the NFW126 (\SFNFWn) morphology, with spectral properties defined above. Finally, we test the evidence for a NFW126 on top of a bulge-GCE diffuse template, and vice versa through nested model comparison.

Errors are estimated from computing an approximated covariance matrix as the inverse of the Fisher information 
matrix from mock data of the best-fit model. See details in Ref.~\cite{Storm:2017arh}.

We notice that the scope of this work is the robust determination of the IG source-count above 10~GeV.
To this end, as demonstrated in \paperI, we do not need to achieve a perfect description of the gamma-ray sky, 
rather to reduce, effectively, the fit residuals. 
\SF~can satisfactorily reach this goal, by optimizing in a data-driven way the gamma-ray diffuse model.
While a better implementation of the Galactic emission is indeed possible (see e.g.~\cite{Pohl:2022nnd}),
deriving the best Galactic 
diffuse gamma-ray model is beyond our scope. We therefore refrain from making model comparison statements, which are in any case prevented 
by the limitations of the maximum likelihood approach.

\subsection{Photon count statistics technique}
\label{sect:dNdS_intro}
Our goal is to measure the flux distribution of point sources in the IG at high energies, and compare their density in different sky regions.
The \Opp~method~\cite{Zechlin:1,Zechlin:2,Zechlin:3} measures this quantity by a statistical analysis of the probability distribution $p_k^{(p)}$ of the photon counts $k^{(p)}$ in each pixel $p$ of a pixelized map. 
The core of the method is based on the fact that different classes of photon sources are expected to contribute to the $p_k^{(p)}$ with different statistics.
The $p_k^{(p)}$ of truly diffuse, isotropic emissions follows a Poissonian distribution. 
Non-Poissonian contributions from point sources and other complex diffuse structures can alter the  probability distribution $p_k$, permitting to investigate these components through the \Opp~of the observed gamma rays.

By exploiting the statistics of photon counts in the data, the \Opp~ extracts the \dnds~ in relation to the number of $k$-photon sources $x^{(p)}_k$ in each pixel, see ~\cite{Zechlin:1} for the details of the mathematical formulation.
The population of point sources is modeled through a \dnds~shaped as a multiple broken power law (MBPL): 
\begin{equation}\label{eq:mbpl}
\frac{\mathrm{d}N}{\mathrm{d}S} = A_{\rm S} \cdot 
\begin{cases} 
\left( \frac{S}{S_0} \right)^{-n_1} \;\;\;\; S > S_{\mathrm{b}1}  \, ;\\
\left( \frac{S_{\mathrm{b}1}}{S_0} \right)^{-n_1+n_2} \left( \frac{S}{S_0} \right)^{-n_2}  \;\;\;\; S_{\mathrm{b}2} < S \leq S_{\mathrm{b}1} \, ;
\\
\vdotswithin{\left(\frac{S_{\mathrm{b}1}}{S_0}\right)} & 
\\
\left( \frac{S_{\mathrm{b}1}}{S_0} \right)^{-n_1+n_2} \left( \frac{S_{\mathrm{b}2}}{S_0} \right)^{-n_2+n_3} \cdots \ \left( \frac{S}{S_0} \right)^{-n_{N_\mathrm{b}+1}}  \\  \hspace{3.8cm} S \leq S_{\mathrm{b}N_\mathrm{b}},\\
\end{cases}
\end{equation}
where the free parameters are the normalization $A_{\rm S}$, the position of the flux break, the indices $n_i$ of the broken power law, and we fix $S_0=5\cdot 10^{-9}$ \fluxunits. We utilize a MBPL with two free breaks $N_b=2$, by sampling directly the parameters in Eq.~\eqref{eq:mbpl}. 
The minimal choice of two breaks is based on our previous results in \cite{Calore:2021jvg} (see Supplemental Material there). Here, as described in Appendix B, we explore the option of a further node at fluxes below $10^{-12}$ ph cm$^{-2}$ s$^{-1}$, finding results compatible with the MBPL fit with 2 breaks down to the methods’s sensitivity. 

Since we do not mask the bright sources, the measured \dnds~ is expected to follow the one of \textit{Fermi}--LAT detected sources in the bright regime. Typically, the \Opp~measures point sources down to (about one order of magnitude)  lower fluxes~\cite{Zechlin:1,Zechlin:2,Calore:2021jvg}. 
We remind that the \Opp~ method in its current implementation does not allow to include in the statistical fit the spatial distribution of the point source component. Thus, the measured point sources   are by construction \textit{isotropic} in the considered region of interest. In \paperI,  we collected insights on the spatial distribution of point sources by subdividing our ROI in sub-regions, and computing the source density as a function of the position in the sky.
Given the limited statistics of the gamma-ray sky at energies larger than 10~GeV, we anticipate that there would not be enough photons to constrain multiple source-count distributions in different sub-regions. 

Similarly to \paperI, we consider within the \Opp-fit the following model components for the gamma-ray sky: (i) A population of isotropic point sources, characterized by their source-count distribution, (ii) an isotropic diffuse emission (with free normalization), (iii)
the Galactic diffuse emission template (as derived by \SF~or from other existing models), 
and (iv) a GCE smooth template as derived by \SF~ and following the bulge or DM morphology, both with free normalization. We give more details on each component in what follows. 
The number of counts in each map pixel $x^{(p)}_\mathrm{diff}$  coming from diffuse templates (ii)-(iv) is defined as: 
\begin{equation}\label{eq:xdiff}
  x_\mathrm{diff}^{(p)} = A_\mathrm{gal} x_\mathrm{gal}^{(p)} +
  A_\mathrm{GCE} x_\mathrm{GCE}^{(p)} + \frac{x_\mathrm{iso}^{(p)}}{F_\mathrm{iso}} F_\mathrm{iso}\,. 
\end{equation}
The quantity $F_\mathrm{iso}$ represents the integral flux of the isotropic diffuse emission, which we use directly  as a sampling parameter in the fit.
The other diffuse templates describe the Galactic diffuse emission and a GCE smooth template. 
As for the GCE smooth template, they correspond to the best fits obtained within each \SF~run (see Sec.~\ref{sec:SF_intro}) for the bulge and DM components. We allow for a rescaling factor $A_\mathrm{GCE}$ relative to the best-fit normalization found by \SF.
Similarly, when we employ the Galactic diffuse emission templates as optimized by \SF, we allow for a $A_\mathrm{gal}$ normalization factor with respect to the best fit. 
To demonstrate the effect of diffuse emission model systematics, we employ other widely used templates, also entering with a $A_\mathrm{gal}$ normalization factor.  Specifically, the official spatial and spectral template released by the \textit{Fermi}--LAT Collaboration for \textit{Pass~8} data (Official P8) ({\tt gll\_iem\_v06.fits}~\cite{2016ApJS..223...26A}), and the model labeled A (modA)  optimized for the study of the DGRB in Ref.~\cite{2015ApJ...799...86A} are tested. We note that in modA   the \textit{Fermi}~bubbles are not modeled.

We fit the gamma-ray sky using the pixel-dependent likelihood function $\mathcal{L}({\bf \Theta})$ as defined within the L2 method in Ref.~\cite{Zechlin:1}, thus taking into account the spatial morphology of the diffuse templates. 
The free parameters ${\bf \Theta}$, together with their prior intervals  are summarized in the appendix~\ref{app:opp_sys}. 
The posterior distribution is defined as $P({\bf \Theta}) = \mathcal{L}({\bf
  \Theta}) \pi({\bf \Theta}) / \mathcal{Z}$, where $\pi({\bf \Theta})$
is the prior and $\mathcal{Z}$ is the Bayesian
evidence. The $P({\bf \Theta})$ is sampled using \texttt{MultiNest}  \cite{2009MNRAS.398.1601F} in its standard configuration, setting 800 live points with a tolerance criterion of 0.2. 
To get prior-independent frequentist maximum likelihood parameter
estimates~\cite{2005NIMPA.551..493R}, we build one-dimensional profile
likelihood functions for each parameter using the final posterior sample.
To perform Bayesian model comparison, the Bayes factors are computed by using the nested sampling global log-evidence  $\ln(\mathcal{Z})$ provided by \texttt{MultiNest}. 

\subsection{Combining SkyFACT and photon-count statistics insights}
\label{sec:combination}

Our hybrid approach, already presented in \paperI, relies on a two-step procedure. First, 
we perform an adaptive-constrained template fit of the gamma-ray IG with \SF.
Second, for each \SF\ run, the corresponding \SF-optimized output models are used as input for the \Opp~ fit. By using the Galactic diffuse emission models as optimized through a \SF~ fit, in \paperI, indeed, we demonstrated that the residuals are reduced and the results of the \Opp~are more stable against diffuse mismodeling. 
We adopt the \SF-optimized components for the Galactic emission, and also for the GCE template, modeled either as the bulge or NFW126 model. 
We then proceed with the \Opp\ fit to \textit{Fermi}--LAT data using the different \SF-optimized Galactic diffuse emission models, the GCE diffuse templates, consistently for each case. We extract the Bayesian evidence $\mathcal{Z}$ and compute the Bayes factors between each model  $i$ and $j$ as
$B_{ij} = \exp(\ln \mathcal{Z}_i - \ln \mathcal{Z}_j)$, since the compared models are not nested anymore. 

The interpretation of the results, and specifically the statistical statements about the GCE morphology  are thus based on (i) the nested model comparison done within the \SF~optimization and (ii) the Bayesian model comparison done fitting again the \SF-optimized templates with the \Opp~method, which adds the modeling of faint point sources. 
Furthermore, the measure of the \dnds~ is extracted within the main ROI, and its characteristics compared to the source count distribution found at high latitudes and in the outer Galactic disk.

\section{\textit{Fermi}--LAT data}\label{sec:data}
Consistently with the dataset analyzed in \paperI, we consider 639 weeks of  \textit{Fermi}--LAT data~\cite{fermidata} until 2020-08-27 and apply standard quality cuts. 
For the \SF~fit, we employ all \texttt{FRONT+BACK P8R3 ULTRACLEANVETO} events (evtype=3) in order to maximize the statistics against the huge number of free parameters in the fit. Our main region of interest (ROI) is the so-called IG defined as 40$^\circ \times 40^\circ$ around the Galactic center. We binned the map into 0.5$^\circ$ pixels. No mask is applied neither on point sources, nor on the Galactic plane. The full energy range (0.2 -- 500 GeV) is divided into 30 logarithmically-spaced bins. The fit is performed either in the energy range $0.3-300$~GeV (27 bins), similarly to \paperI, or in the high-energy range 10 -- 300 GeV (13 bins).

The benchmark choice for the  \Opp~fit is to consider all \texttt{P8R3 CLEAN} events and  focusing in a single bin covering the 10--300~GeV energy range.
Given the limited number of photons in the considered high-energy range, this choice increases the statistics of photon counts with which the \Opp~can measure the \dnds~down to low fluxes.
Since the point spread function (PSF) of \textit{Fermi}--LAT at these energies is expected to be  of the order of 0.1 -- 0.5$^\circ$, i.e.~smaller or comparable to the pixel size, we consider all PSF quartile events.
The \Opp~performance has been tested using \texttt{CLEAN} \textit{Fermi}--LAT high-energy events elsewhere \cite{DiMauro:2017ing}. Even less stringent quality cuts (\texttt{SOURCE} events) have been used in past analysis of the IG to increase the sensitivity of photon-count statistical methods \cite{Linden:2016rcf}.
Nevertheless, to test the stability of our conclusions against the event selection, we discuss results using \texttt{P8R3 ULTRACLEANVETO} events with and without further selecting the quartile of best angular resolution (PSF3), see appendix~\ref{app:opp_sys}. 
All the components entering the \Opp~fits are corrected for the PSF effect as detailed in Ref.~\cite{Zechlin:1}. 
The photon events are binned spatially using 
 the HEALPix equal-area pixelation scheme~\cite{2005ApJ...622..759G} with resolution parameter $\kappa=7$ \cite{2005ApJ...622..759G}.

Different ROIs in the sky are considered in this work. 
The IG ROI is shared by both the \SF~and \Opp~analysis. 
Cuts at latitudes $|b| >0.5^\circ$ or $ 1^\circ$ are introduced to check the stability of \Opp~results. 
We note that, in an effort to maximize the statistics for the  high-energy fit,  we increased the ROI for the \Opp~analysis with respect to the  $20^\circ \times 20^\circ$ IG region used in \paperI. We tested that the main results are unchanged when reducing the ROI to the one of \paperI.
As in \paperI, two control regions are introduced to compare the source density with respect to the IG: the outer Galaxy (OG, $|b|<20^\circ$, $60^\circ<|l|<90^\circ$, cutting the inner $|b| <1^\circ$) and the extragalactic region (EG, $|b|>40^\circ$, $|l|>90^\circ$).
Figure~\ref{fig:counts} shows the \textit{Fermi}--LAT data counts map for the IG, OG, EG ROI (left to right) in the energy bin 10--300~GeV. The total number of counts within each region for \texttt{CLEAN} events is 92529 (IG), 50619 (OG) and 52324 (EG). 

\section{Results: excess significance and morphology with skyFACT}
\label{sec:SF_results}
In this section, we discuss the results of the fits to the gamma-ray sky with \SF\ with the aim of determining the excess significance and its morphology at high energies. 

First, we run the fit in the OG ROI in the 10 -- 300~GeV energy range to serve as a control region.
We check the evidence for an additional GCE component on top on the \texttt{SF-noGCE} model.
For the GCE, we consider either an NFW126 or a Bulge template. 
In none of the two cases, we find significant evidence for the GCE ($\sigma \sim 0$). This
result is consistent with the fit in the full energy range, 0.3 -- 300 GeV, performed in \paperI.
Therefore, the baseline \SF\ model in the OG results from the optimization of the sky components without GCE and it will be used as input for the subsequent \Opp\ fit.

\begin{figure}[t]
\includegraphics[width=0.48\textwidth]{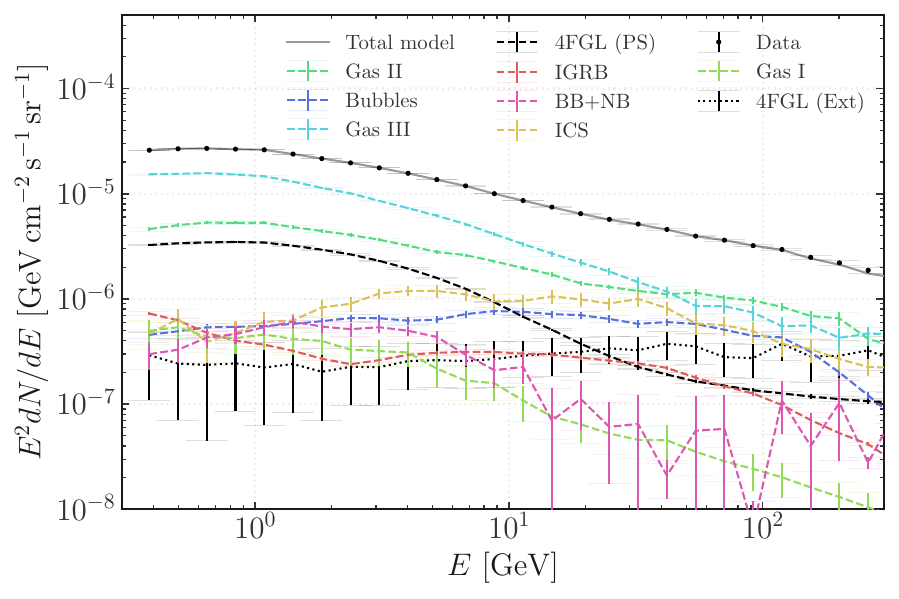}
\caption{\label{fig:skyfact_IG_fit} 
{\it \SF\ best-fit spectral components}. Results of the IG analysis in the 0.3 -- 300~GeV energy range,
for the preferred \texttt{SF-B} model composed by the background model, and an additional GCE Bulge component (i.e.~BB+NB). 
See section~\ref{sec:SF_intro} for more details on the
model components.  
}
\end{figure}

Second, we analyze the IG ROI. 
We first consider the full energy range, 0.3 -- 300 GeV, and test the evidence for an additional GCE component, as explained in 
section~\ref{sec:SF_intro}.
The GCE spatial template is fixed and assumed to follow either NFW126 or Bulge.
The input spectrum is a power law with index -2.5, and it is allowed to freely vary energy bin by energy bin.
No matter what GCE template is considered, we found strong evidence for this additional component:
$\sigma \sim 13$ (9.9) for the additional Bulge (NFW126) template.
We then check the preference for Bulge on top of NFW126 and, viceversa, of NFW126 on top of Bulge.
Analogously to  \paperI, we find that it is necessary to add Bulge on top of NFW126 ($\sigma =$ 9.1), 
while there is no need for an additional NFW126 on top of the Bulge ($\sigma < 4$).
We notice that in the full energy range and adopting a power-law input spectrum for the GCE, 
the evidence for the GCE and the preference of the Bulge on top of NFW126 is consistent 
with the results found in \paperI\ for a MSP-like input spectrum, i.e.~a power law (index -1.46) 
with exponential cutoff at 3.6~GeV, see also discussion in appendix~\ref{app:SF_sys}.
We can understand these results by considering that, in the full energy range, the fit is driven by the low-energy data (1 -- 5~GeV). 
Therefore the possible high-energy tail of the excess, if there, does not strongly weight in determining 
the evidence for the GCE and Bulge.
In figure~\ref{fig:skyfact_IG_fit}, we report the best-fit spectra of the \texttt{SF-B} run (i.e.~including the Bulge GCE) 
in the full energy range 0.3 -- 300 GeV.
The best-fit spectrum for the Bulge (run with power-law input spectrum) is consistent with the one in \paperI~(run using an input MSP-like spectrum for the Bulge) for energies below 10~GeV,
while at higher energies the Bulge shows significant emission, see also discussion in appendix~\ref{app:SF_sys}.

Since the best-fit results in the full energy range are driven by the low-energy data, in order to properly test the presence of the GCE
at higher energies we need to run the fit in a narrower energy range and evaluate the significance of 
additional components therein. 
We therefore consider the 10 -- 300 GeV energy range with  same setup as above for the different sky components.
We summarize the results in table~\ref{tab:SF_HE_fit}: The first result is that we find strong evidence for the GCE (either 
modeled with a Bulge or an NFW126 template) also at high energies, $\sigma > 5$ regardless of the GCE spatial template. 
Secondly, our fit prefers the Bulge model over the NFW126 at high significance ($\sigma = 5.5$). While the evidence is lower than the one found in the full energy range, it still
strongly indicates the need of the GCE when fitting the high-energy {\it Fermi} sky.
Similarly, at high energies, there is no evidence for an additional NFW126 template on top of the Bulge component ($\sigma = 0.8$).
The integrated flux of the Bulge component in this case is about a factor of two higher than the best-fit integrated flux of the Bulge above 10 GeV from the fit in the full energy range.
We will demonstrate that this difference does not impact the \Opp\ results which are stable against variation of the overall
Bulge input intensity.
In appendix~\ref{app:SF_sys}, we discuss further systematics that may affect the evidence of the GCE in the high-energy interval, and show that this evidence is robust against them.
Specifically, the influence on the results of the modeling of the  {\it Fermi} bubbles  is scrutinized.
We find a strong evidence for the GCE at high energies in all the model variations inspected. We stress that all our  {\it Fermi} bubbles templates possess a low-latitude component. We therefore find strong evidence for the GCE at high energy {\it in the presence} of the
{\it Fermi} bubbles at low latitudes.

\begin{figure*}[t]
\includegraphics[width=0.326\textwidth]{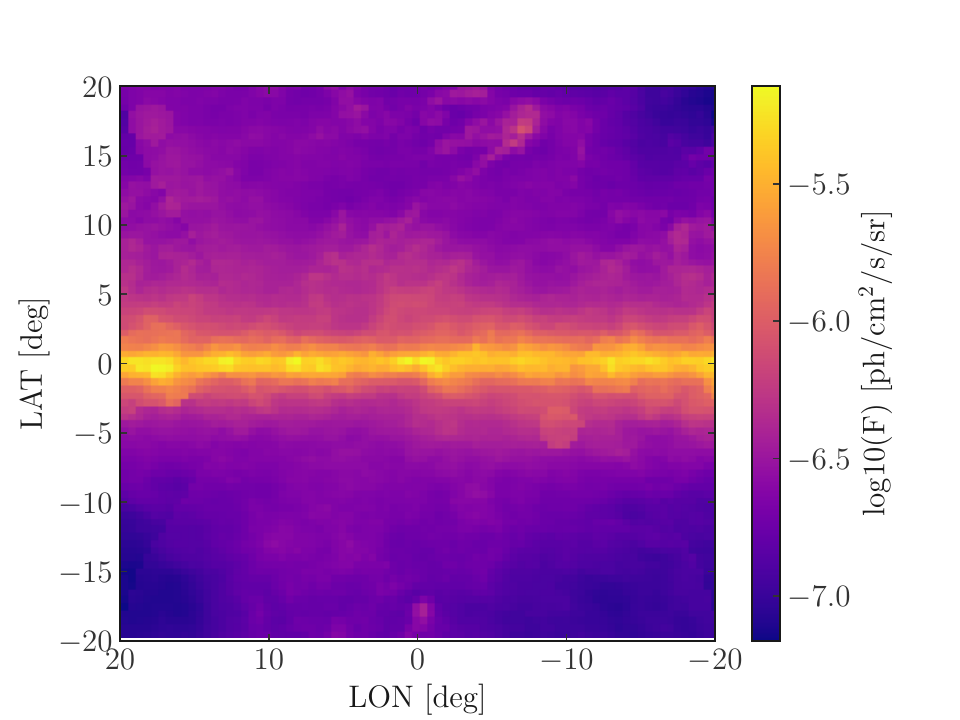}
\includegraphics[width=0.326\textwidth]{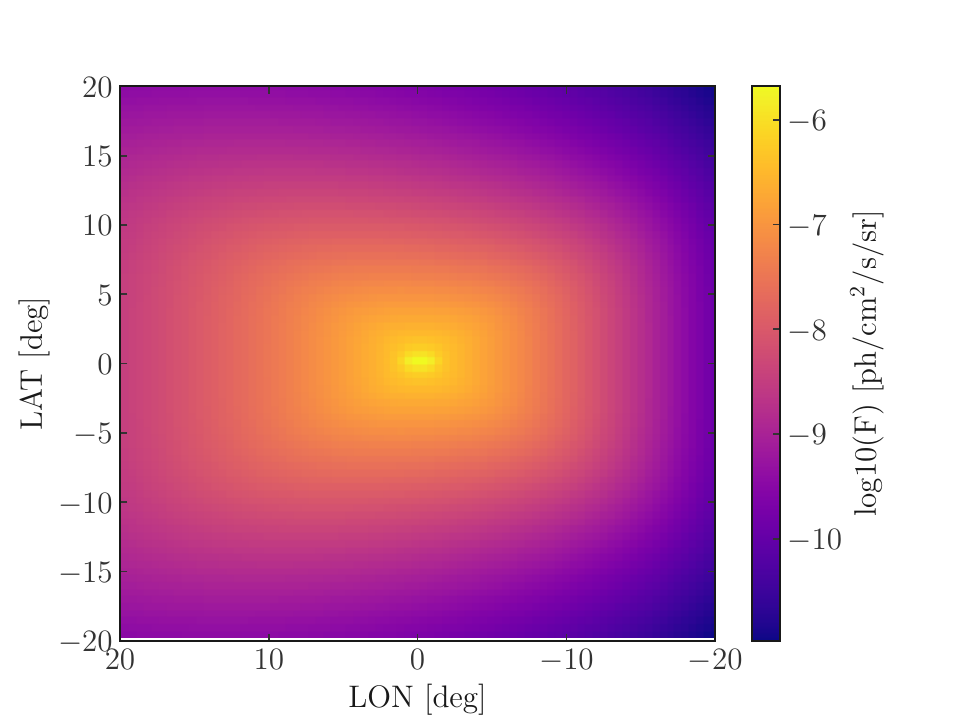}
\includegraphics[width=0.326\textwidth]{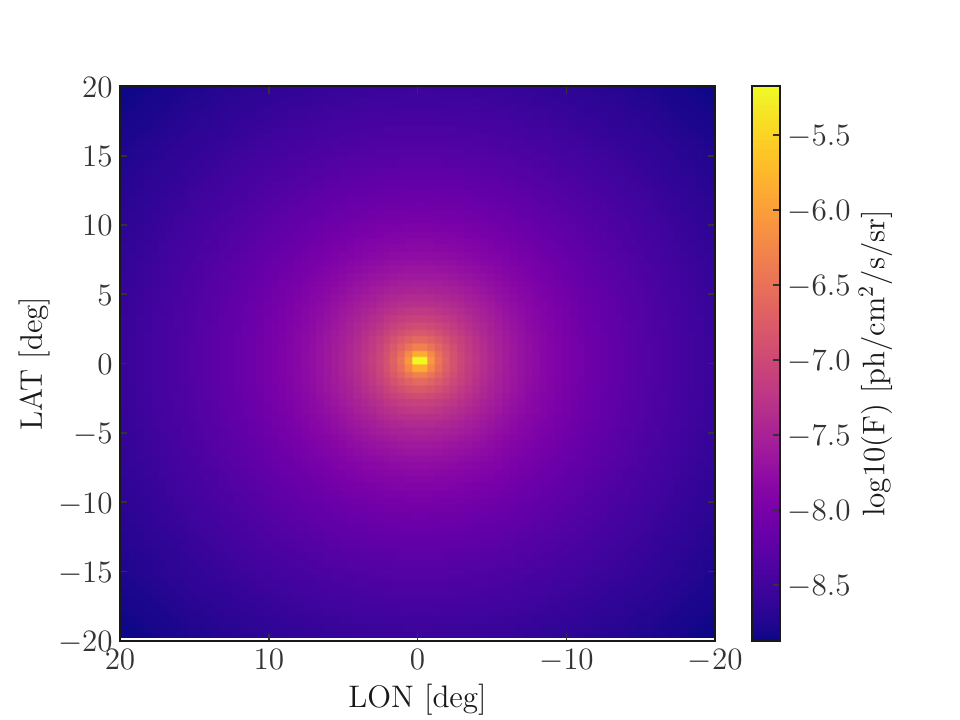}
\caption{\label{fig:model_maps} \textit{The \SF-optimized templates} for the Galactic diffuse emission (left), the sum of nuclear and boxy stellar bulge (the Bulge, center), and NFW126 dark matter as integrated in the 10 -- 300~GeV energy bin. Note the different scales in each panel. }
\end{figure*}

We remind the reader that our scope is not to find the overall best-fit gamma-ray sky model, rather to provide an 
input model for the \Opp\ analysis which maximally reduces residuals in an effective way and allows us to derive
robust conclusions on the source-count distribution. 
The \SF\ fit results for the \texttt{SF-B} model in the 10 -- 300 GeV energy interval will be used as input for the \Opp\ analysis in section~\ref{sec:Opp_results}.
The baseline templates are shown in figure~\ref{fig:model_maps}.

\begin{table*}
\caption{{\it Excess evidence in the IG.} Results of the \SF\ gamma-ray fit in the IG in the high-energy interval, 
showing the evidence (in units of $\sigma$) of the GCE itself, and of the Bulge and NFW126 templates on top of the baseline background model. In all runs, the GCE morphology
is fixed, while the spectrum is allowed to freely vary energy bin by energy bin. The input GCE spectrum is a power law with index -2.5.}
\centering
\begin{tabular}{ l @{\hspace{10px}}|  c @{\hspace{10px}}|c @{\hspace{10px}} |c  @{\hspace{10px}} } \hline\hline
& GCE Bulge (NFW126)	&  Bulge over NFW126  & NFW126 over Bulge \\ \hline
 {\bf Evidence} ($\sigma$) & 8.1 (5.6) & 5.5 & 0.8   \\ 
\hline\hline
\end{tabular}
\label{tab:SF_HE_fit}
\end{table*}

\section{Results: source--count distribution}
\label{sec:Opp_results}
In this section we discuss the results for the source--count distribution \dnds~ as obtained analyzing \textit{Fermi}--LAT data with the \Opp.

\begin{figure*}[t]
\includegraphics[width=0.48\textwidth]{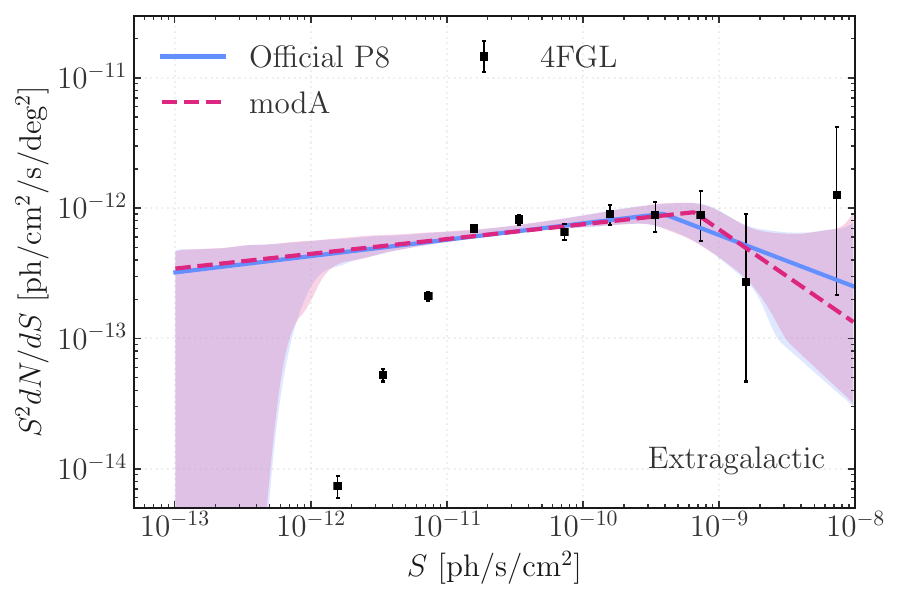}
\includegraphics[width=0.48\textwidth]{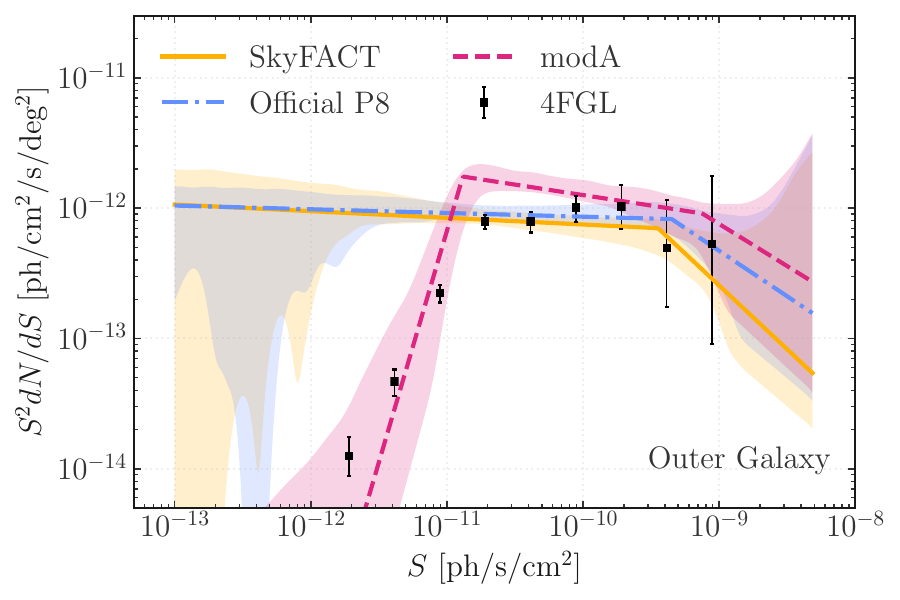}
\caption{\label{fig:EG_OG} 
\textit{Extragalactic and Outer Galaxy \dnds.} Source-count distribution of the EG ROI (left panel) and  of the OG (right panel) as obtained from the \Opp~analysis. 
 The \dnds~result using the Official P8 (blue lines) model, modA (magenta lines) and the \SF~Galactic diffuse emission model without any component modeling the GCE (yellow line, only right panel) are reported. 
The $1\sigma$ uncertainty bands are indicated with colored areas. 
The count distribution of 4FGL sources is illustrated with black points for comparison in the bright flux regime. }
\end{figure*}

\subsection{Extragalactic and Outer Galaxy sky regions}
The result for the \dnds~ of the EG (OG) sky is 
illustrated in the left (right)  panel of figure~\ref{fig:EG_OG}. 
We report the result obtained with the \Opp~analysis of CLEAN all PSF events in the 10--300~GeV energy range when using the Official Pass8 diffuse emission template (blue) and the modA (magenta). 
As for the OG ROI, we also show the results obtained using an optimized \SF~ diffuse template (yellow) without including a GCE template. 
The best fit results are reported together with the $1\sigma$ uncertainty bands. 
The source-count distribution  from the sources in the 4FGL is shown for reference with black points, where uncertainties include statistical uncertainties on the source count only. 
As expected, the \Opp~recovers the \dnds~of bright sources in agreement with the 4FGL number counts down to fluxes of about $S\sim 10^{-11}$~\fluxunits.
For lower fluxes, the \dnds~of 4FGL sources decreases quickly, while the \Opp~ measures the collective contribution of faint point sources down to at least one order of magnitude at  $S\sim 10^{-12}$~\fluxunits.
At even lower fluxes, the $1\sigma$ uncertainty band increases significantly, and the measured \dnds~ becomes fully degenerate with the diffuse isotropic emission (in other words, the position of the second break of the \dnds~ and the normalization of the diffuse isotropic emission are highly correlated). 
This measurement of the \dnds~ is robust against possible systematics coming from the \dnds~ parametrisation as MBPL or using the Hybrid approach, as well as the event selection, see the appendix~\ref{app:opp_sys}. 

In the EG ROI, consistent \dnds~ are measured  when using the Official Pass8 and the modA diffuse models, confirming the robustness of the results against diffuse emission mismodeling at high latitudes, similarly to what observed in the 2--5~GeV energy range in \paperI. 
The measured \dnds~ is in overall agreement with previous \Opp~results obtained within similar event selection, energy bin and region of interest in the high latitude sky~\cite{Zechlin:2}. 
As expected, it becomes unconstrained below about $ 6 \cdot 10^{-13}$~\fluxunits.

Using the Official P8 diffuse model and the \SF~ optimized one, the \dnds~ of the OG is found to be well described by a power law of index $n_2=2.05^{+0.17} _{-0.08}$, from $5 \cdot 10^{-10}$ \fluxunits\ down to $5 \cdot 10^{-12}$ \fluxunits~, where the uncertainty band opens by few orders of magnitude. 
These \dnds~ results are compatible with the source counts of bright sources within the 4FGL catalog, and reveal unresolved sources down to at least $5 \cdot 10^{-12}$ \fluxunits. 
Instead, we observe  effects of mismodeling the diffuse emission when using  modA, while the results using the \SF~and the Official P8 model are consistent within statistical uncertainties. 
Similarly to what observed in \paperI, in this scenario the \Opp~looses its sensitivity and fails to measure faint sources at fluxes lower than about $10^{-11}$ \fluxunits. 
The \dnds~measured by the \Opp~with modA suggests that residuals in the fitted gamma-ray sky are interpreted as bright point sources around the sensitivity threshold of the 4FGL catalog. 
The measure of the \dnds~of gamma-ray sources in the OG ROI  in the 10--300~GeV energy range 
is a novel result of this work.

\begin{figure*}[t]
\includegraphics[width=0.48\textwidth]{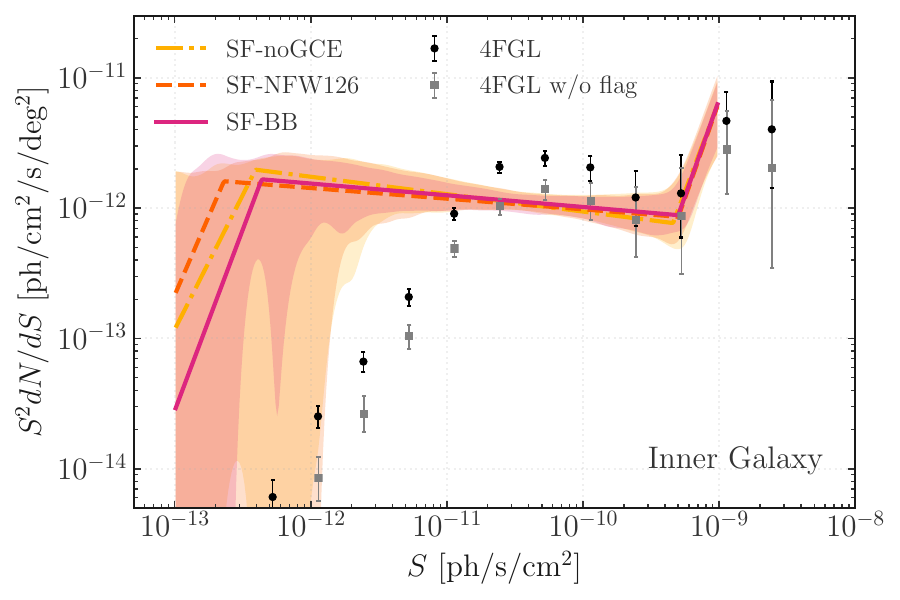}
\includegraphics[width=0.48\textwidth]{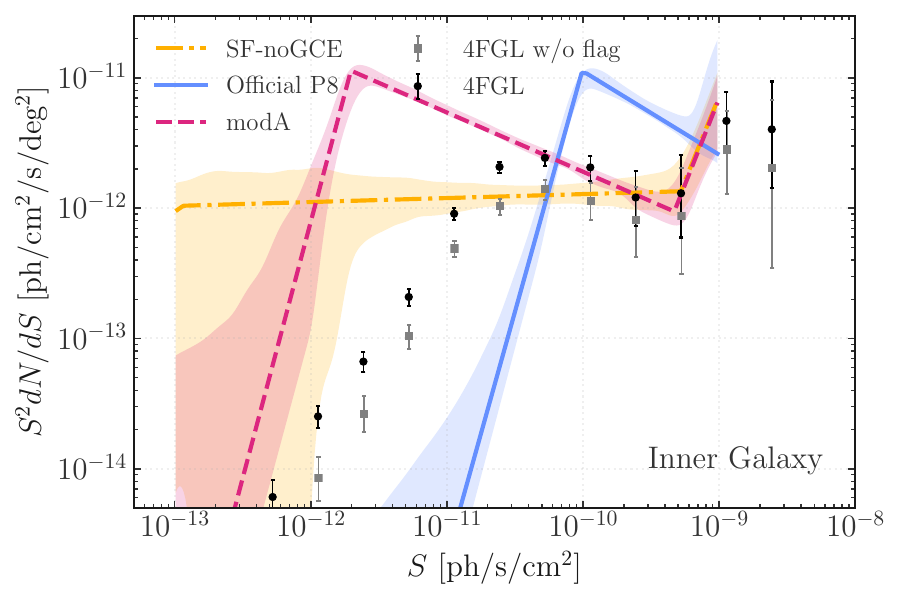}
\caption{\label{fig:IG} 
\textit{Inner Galaxy \dnds.} Source-count distribution of the IG ROI measured by the \Opp~using different \SF-optimized diffuse models (left panel, for $|b|>1^\circ$) and compared to non-optimized models from the literature (right panel, for $|b|>0.5^\circ$).
The $1\sigma$ uncertainty bands are indicated with colored areas. 
The count distribution of 4FGL sources is illustrated with black points for comparison in the bright flux regime. Gray points represent the count distribution of 4FGL sources without any analysis flag, see \cite{Fermi-LAT:2019yla}.}
\end{figure*}

\subsection{Inner Galaxy}
We proceed to measure the source count distribution of bright and faint point sources within the IG ROI in the 10--300~GeV energy bin, and considering different Galactic diffuse emission models and GCE diffuse models as obtained in the dedicated \SF~fit described in Sec.~\ref{sec:methods}. 
When using the \SF-optimized diffuse models, we consistently include in the \Opp~fit the GCE diffuse model following a NFW126 or a bulge morphology, with a free overall normalization, apart for the \texttt{noGCE} setup. This permits us to evaluate again, this time with Bayesian statistics as detailed in Sec.~\ref{sec:combination}, which model is preferred when fitting the IG including also faint point sources. 
The results for the \Opp~fits using the diffuse models obtained within the \texttt{SF-noGCE}, \texttt{SF-NFW126} and \texttt{SF-B} fits are summarised in Table~\ref{tab:results} for the \Opp~fits of CLEAN allPSF data. The columns illustrate the latitude cut, the Bayesian log-evidence, the  point source fluxes for $S<10^{-10}$ \fluxunits~ and the normalization of the diffuse GCE for the different configuration explored. 
Regardless of the latitude cut, we find that the model best describing the gamma ray data is the one in which the Galactic diffuse emission corresponds to the \SF~fit and that also include a smooth, diffuse GCE following a bulge morphology with Bayes factor $\ln B_{ij} \gtrsim 30$. 
Adding a diffuse GCE template is also preferred with respect to the \texttt{SF-noGCE} case with similar Bayes factor.
Both the Galactic diffuse emission and the GCE smooth templates are found in the \Opp~fit with normalizations consistent with one. 
We tested that similar results are obtained using UCV allPSF and PSF3 data selections.
This corroborates the result obtained within \SF: The IG gamma-ray sky at energies larger than 10~GeV is best described when a smooth GCE following a bulge morphology is included in the model, also when modeling the faint point sources. 

The PS flux is found to be consistent among the different optimized Galactic diffuse models employed, and slightly higher when cutting only the inner $0.5^\circ$, see next section for further discussion on the source density. By integrating all the PS fluxes below $S<10^{-10}$ \fluxunits~ (and thus in an intermediate regime between bright and faint), we obtain a total PS flux which is comparable to the total flux measured by \SF~  in the full IG ROI within the fit for $E>10$~GeV, equal to $2.37\cdot 10^{-8}$~\fluxunits. 

\begin{table*}
\caption{Results for the \Opp~analysis of the IG LAT data, CLEAN, all PSF. First three columns:  setup of the analysis and  latitude mask of the IG. The $\ln(\mathcal{Z})$ is the nested sampling global log-evidence extracted from \texttt{Multinest}.  Last two columns:  flux from  IG point sources (in units of 
$10^{-8}$ \fluxunits, and for $S<10^{-10}$ \fluxunits), and  normalization of smooth GCE template in the \Opp~fit when relevant.}
\centering
\begin{tabular}{ l @{\hspace{10px}}|  c @{\hspace{10px}}|c @{\hspace{10px}} |c  @{\hspace{10px}}| c} \hline\hline

\textbf{Description } & $|b|$ cut [$^\circ$]& $\ln(\mathcal{Z})$ 	&   PS flux   & $A_{\rm B/NFW126}$ \\ \hline

 No GCE   & 0.5	& $ -19056.81 $ 	& $3.42^{+0.78}_{-1.58}$ & - \\ 

 NFW126  &  	0.5	& $ -19060.14 $ 	& $3.76^{+0.41}_{-1.75}$ & $0.84^{+0.10}_{-0.16}$ \\ 

 Bulge   &    0.5	& $ -19027.22$ 	& $3.94^{+3.89}_{-1.98}$ & $0.97^{+0.06}_{-0.08}$  \\ 

    \hline

  No GCE   &   1	& $-18085.2 $ 	& $2.75^{+0.76}_{-1.39}$ & - \\ 

 NFW126  &   1	& $ -18072.6 $ 	& $2.84^{+0.7}_{-1.3}$ & $1.45^{+0.05}_{-0.15}$ \\ 

 Bulge   &   1	& $ -18054.34$ 	& $2.68^{+1.07}_{-1.08}$ & $1.02^{+0.08}_{-0.04}$ \\ 

 \hline\hline
\end{tabular}
\label{tab:results}
\end{table*}

In addition, we find that a consistent population of faint point sources is reconstructed  no matter the \SF-optimized Galactic diffuse emission model. This is illustrated in the left panel of figure~\ref{fig:IG}.
The \dnds~as measured by the \Opp~ and cutting the inner $|b|<1^\circ$ is reported for the \texttt{SF-noGCE}, \texttt{SF-NFW126}, and \texttt{SF-B} setups with different colors, together with the $1\sigma$ uncertainty band. 
No matter the setup of diffuse templates used, the \Opp~consistently measures faint point sources down at least to $10^{-12}$ \fluxunits. At brighter fluxes and up to $10^{-11}$ \fluxunits, the reconstructed \dnds~is found compatible with the source count in the 4FGL, in particular when flagged sources are excluded. 
Between $10^{-12}$ \fluxunits and $10^{-11}$ \fluxunits the \Opp~measures a number of point sources significantly higher than the ones in the 4FGL.
Between $10^{-9}$ \fluxunits and $10^{-12}$ \fluxunits, the \dnds~ follows a power law with index $n_2\sim 2$. 
At fluxes lower than $10^{-12}$ \fluxunits the uncertainty band opens significantly and the second break in the \dnds~becomes fully degenerate with the isotropic diffuse background normalization. 

In the right panel of figure~\ref{fig:IG} we illustrate the effect of Galactic diffuse emission mis-modeling  for $|b|<0.5^\circ$. 
When using non-optimized Galactic diffuse emission models such as the Official P8 (blue line) or the modA (magenta), the \Opp~method is not able to reconstruct bright point sources even at high fluxes of $10^{-10}$ \fluxunits, and spurious peaks in the \dnds~are found with high significance. 
We attribute this effect to the residuals left when analyzing gamma rays at low latitudes, which are misattributed by the \Opp~to a bright population of point sources.

\subsection{Spatial distribution: Source density}
In order to characterize the spatial distribution of the point source population measured by the \Opp~we compute the source density in the different ROIs analyzed. 
We remind that, in the current implementation, the \Opp~method measures an \textit{isotropic} population of point sources in the region of interest, and thus is not sensitive to the pixel-by-pixel spatial distribution of sources. 
Thus the measured \dnds, and in turn the source density are to be considered as  average quantities within the ROI. 
We  compute the source density by integrating the measured \dnds~(and the corresponding $1\sigma$ uncertainty) in the  flux interval [$3\cdot 10^{-12}, 1\cdot 10^{-11}$] ph cm$^{-2}$ s$^{-1}$. 
This flux interval corresponds to the regime in which the 4FGL is incomplete for the source counts, and thus characterizes the unresolved point sources collectively measured by the \Opp. 
The source density obtained integrating the \dnds~as measured in the EG ROI  is  of $\sim 0.12^{+0.02} _{-0.01}$~sources/deg$^2$. 
The OG source density is found to be slightly higher, at the level of  $0.20 ^{+0.09} _{-0.03}$ sources/deg$^2$, suggesting that Galactic sources contribute to the \dnds~in addition to the extragalactic sources measured in the EG.

The source density in the full IG ROI  is  $0.31 ^{+0.11} _{-0.09}$ sources/deg$^2$ when  cutting the inner $|b|<1$~degree, and $0.28 ^{+0.10} _{-0.09}$ sources/deg$^2$ when  cutting the inner $|b|<0.5$~degree.
Both results are compatible within errors, and reveal a source density higher with respect to the EG and OG. 
 However, the reduced photon statistics in the 10--300~GeV energy bin prevents to robustly measure the radial and longitude profiles of point sources using multiple regions as done in \paperI. 
We thus provide the source density within the North and South hemispheres, corresponding to $0.35 ^{+0.20} _{-0.24}$ sources/deg$^2$ and $0.33 ^{+0.22} _{-0.29}$ sources/deg$^2$
respectively, as well as for positive and negative longitudes, equal to $0.47 ^{+0.30} _{-0.32}$ sources/deg$^2$ and $0.18 ^{+0.15} _{-0.15}$ sources/deg$^2$.
The source density is found to be slightly higher than the OG and the average IG level within positive longitudes. We note that, if the GCE consists of MSPs within the stellar bulge of the Milky Way, and if they contribute to the unresolved point sources in the considered flux interval, an asymmetry among negative and positive longitudes is indeed expected \cite{Bartels_bulgelum,2013MNRAS.434..595C,2020MNRAS.495.3350C}, being the stellar bulge more luminous at positive longitudes, see e.g.  figure~\ref{fig:model_maps}.
However, we leave the interpretation of these measurements to forthcoming work.

\section{Conclusion and Outlook}
\label{sec:conclusions}
\begin{figure}[t]
\centering 
\includegraphics[width=0.52\textwidth]{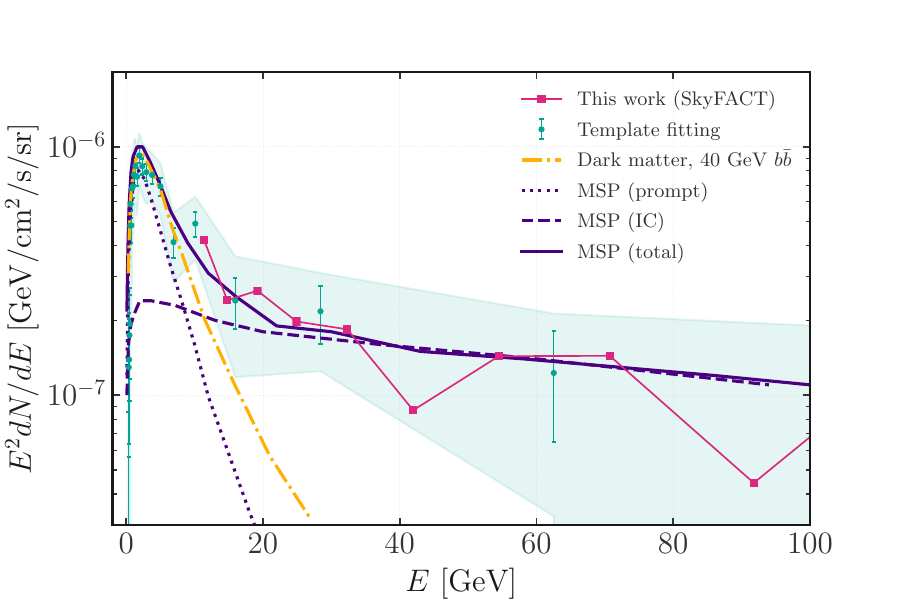}
\caption{\label{fig:GCE_hetail} The GCE energy spectrum is illustrated in linear scale to highlight the $E>10$~GeV tail.  
Our results obtained with \SF\ for energies larger than $10$~GeV are shown in magenta. 
The measured spectra from the template-based analysis of Ref.~\cite{Calore:2014nla}  is reported with blue points, where the shaded band encloses the systematic uncertainties on the GCE spectra when varying the Galactic diffuse emission modeling. 
Model interpretations assuming dark matter annihilations (Ref.~\cite{Cholis:2021rpp}, yellow dot-dashed line, rescaled by a factor 1.5) or MSP prompt (purple dotted) plus IC emission (dashed) from Ref.~\cite{Gautam:2021wqn} are overlaid for comparison.}
\end{figure}
In this work, we presented a new study of the IG \textit{Fermi}--LAT data at energies above 10 GeV 
with the goal of investigating the robustness of the high-energy tail of the GCE, and assessing the role of sub-threshold point sources. 
Our analysis is based on an innovative method which combines 
adaptive template fitting and pixel count statistical methods, while minimizing the mis-modelling of Galactic diffuse emission backgrounds. The present paper extends our previous results of \paperI~focused on the 2-5 GeV energy bin, where the GCE is prominent. 

We performed adaptive-constrained template fits with \SF, firstly in the full energy 0.3 -- 300~GeV, and then in the high-energy range 10 -- 300 GeV. The gamma-ray sky was interpreted as the combination of several components, each one modeled by a specific spectral and morphological input template. In particular, we considered a NFW126 or a Bulge template for the GCE. 
We found a strong evidence for the GCE also at high energies, $\sigma > 5$ regardless of the GCE spatial template. 
Remarkably, our fit preferred the Bulge model over the NFW126 at high significance ($\sigma = 5.5$) and showed no evidence for an additional NFW126 template on top of the Bulge component ($\sigma = 0.8$). 
We  find strong evidence for the
GCE at high energy in the presence of the \textit{Fermi} bubbles
at low latitudes.

The \SF~fit results, which maximally reduce residuals in an effective way, were then employed as input model for the \Opp\ analysis. 
After measuring the \dnds~in the OG and EG control regions, we proceeded to measure the source count distribution of bright and faint point sources within the IG ROI in the 10--300~GeV energy bin, and considered different Galactic diffuse emission models and GCE diffuse models as obtained in the dedicated \SF~fit. 
We found that the model best describing the gamma-ray data 
is the one that contains the \SF~fit for the Galactic diffuse emission {\em with } the inclusion of a smooth, diffuse GCE following a bulge morphology. 
Adding a diffuse GCE template is also preferred with respect to the \texttt{SF-noGCE} case.
Both the Galactic diffuse emission and the GCE smooth templates are found in the \Opp~fit with normalizations consistent with one. 
Notwithstanding the low statistics characterizing the emission at energies $> 10$ GeV, the IG gamma-ray sky at energies larger than 10~GeV is best described when a smooth GCE following a bulge morphology is included in the model, also when modeling the faint point sources. 
A further result reached by the \Opp~is that a consistent population of faint point sources is reconstructed. No matter the \SF-optimized Galactic diffuse emission model, the \Opp~consistently measures faint point sources down at least to $10^{-12}$ \fluxunits. Between $10^{-12}$ \fluxunits and $10^{-11}$ \fluxunits the \Opp~measures a number of point sources significantly higher than the ones in the 4FGL.

In the current implementation, the \Opp~method is not sensitive to the pixel-by-pixel spatial distribution of sources. In an attempt to  
characterize the spatial distribution of the point-source population measured by the \Opp, we computed the source density in the EG, OG and IG ROIs by integrating the measured \dnds~in the  flux interval [$3\cdot 10^{-12}, 1\cdot 10^{-11}$] ph cm$^{-2}$ s$^{-1}$. We found an IG source density higher with respect to the EG and OG. 
Similar source densities are found within the North and South hemispheres, 
while a higher density was found for positive longitudes with respect to negative ones. If GCE were explained by point sources such as MSPs, one would expect an asymmetry among negative and positive longitudes, being the stellar bulge indeed more luminous at positive longitudes.

The spectrum of the GCE as found in our \SF~ fit to the \textit{Fermi}--LAT data at energies above 10 GeV is reported in figure~\ref{fig:GCE_hetail}, along with earlier results. 
The linear $x$-axis emphasizes the high-energy tail of the GCE, 
where we  
confirm and support the existence of a significant high-energy tail in the GCE spectrum at energies of $E>10$~GeV, consistently with the results from template fitting by e.g. Refs.~\cite{Calore:2014nla}.  
For illustrative purpose, we overlay two model interpretations picked from the literature. 
Dark matter annihilation of weakly interacting massive particles  with $m_{\rm DM}=40$~GeV in the $b \bar{b}$ channel ~\cite{Cholis:2021rpp} (yellow dot-dashed line, rescaled by a factor 1.5 to match the normalization of GCE spectrum found in Ref.\cite{Calore:2014nla}) can explain the peak of the GCE at around few GeV, but fails to explain the high-energy tail.  
Annihilation into a combination of different final states as suggested by specific dark matter models could explain the spectrum until few tens of GeV, but they are subject by other constraints, see e.g. \cite{DiMauro:2021qcf,Calore:2022stf,Balan:2023lwg,Foster:2022nva}. 
Conversely, the cumulative flux from MSPs in the Galactic bulge, including the prompt (purple dotted) and inverse Compton (IC, dashed) emissions as modeled by Ref.~\cite{Gautam:2021wqn} have a spectrum compatible with  both the GCE peak and the high-energy tail, depending on the spectral properties of the $e^\pm$ population emitted~\cite{Petrovic:2014xra}. 

The robustness of our results confirming the existence of a GCE at high energies, demonstrating the preference of the \textit{Fermi}--LAT data for a Bulge morphology and measuring a population of point sources below the detector threshold, motivates further studies towards the understanding of the GCE in terms of point sources and specifically of MSPs. This exploration, along with other viable and complementary interpretations, are left for future study.

\begin{acknowledgments}
We warmly acknowledge C. Eckner for discussion on \SF\ performance and careful reading of the manuscript, and D. Malyshev and N. Rodd for careful reading of the manuscript and useful comments. 
SM acknowledges the European Union's Horizon Europe research and innovation program for support under the Marie Sklodowska-Curie Action HE MSCA PF–2021,  grant agreement No.10106280, project \textit{VerSi}. FC acknowledges support by the ``Agence Nationale de la Recherche'', grant n.~ANR-19-CE31-0005-01. F.D. acknowledges the support of the Research grant {\sc TAsP} (Theoretical Astroparticle Physics) funded by Istituto Nazionale di Fisica Nucleare. 
\end{acknowledgments}

\bibliography{biblio}

\newpage 

\renewcommand{\thetable}{A\Roman{table}}  
\renewcommand{\thefigure}{A\arabic{figure}}
\renewcommand{\theequation}{A\arabic{equation}}

\appendix
\section{\SF\ fits: systematics and fit ranges}
\label{app:SF_sys}
In this section, we test our \SF\ fit results against a number of systematics.

\begin{figure*}[t]
\includegraphics[width=0.48\textwidth]{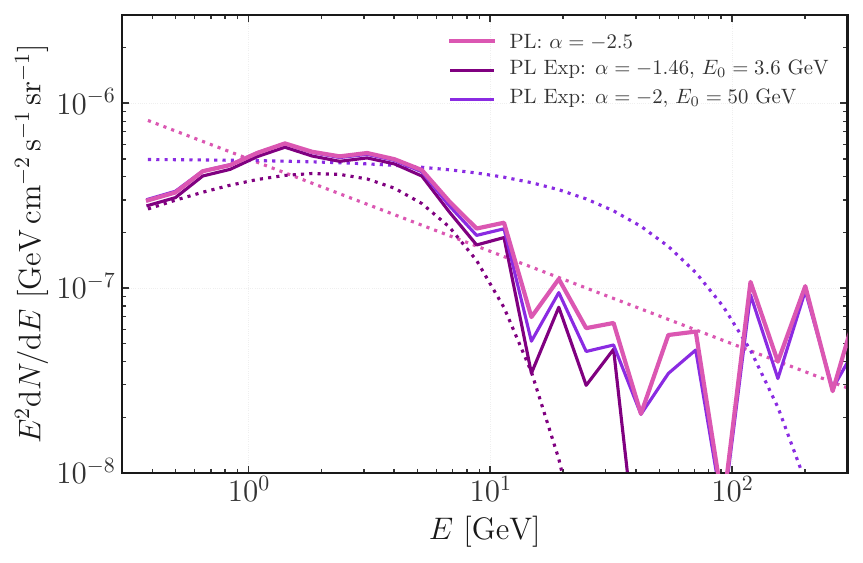}
\includegraphics[width=0.48\textwidth]{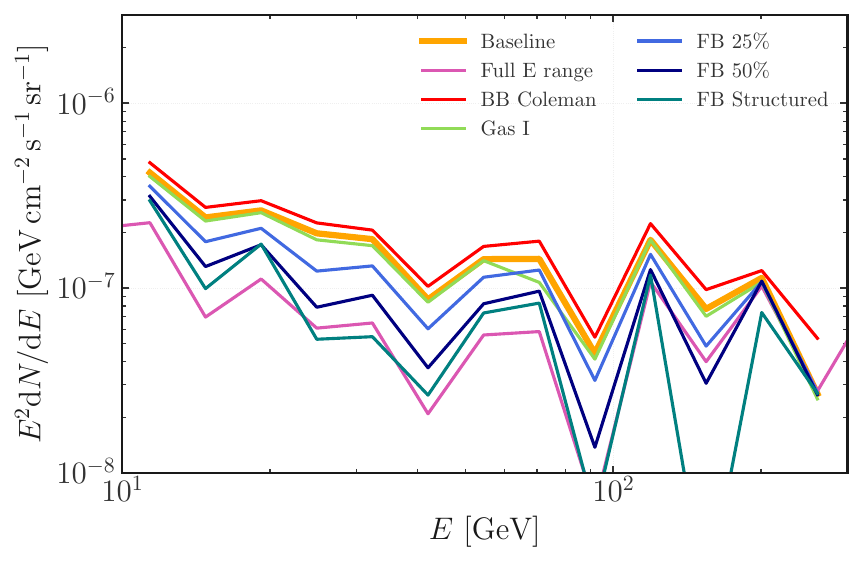}
\caption{\label{fig:SF_spectra_syst} \textit{Systematic tests on GCE spectrum.} Left: GCE best-fit spectra  (BB+NB components) reconstructed with the \SF\ fit in the range 0.3 -- 300 GeV, adopting different input spectra: a power law with index -2.5 (deep pink solid, baseline choice), a power law with exponential cutoff with parameters $\alpha = -1.46$ and $E_0 = 3.6$ GeV (purple, typical MSP-like ``prompt'' spectrum), or $\alpha = -2$ and $E_0 = 50$ GeV (violet). Dotted lines correspond to the input spectra.
Right: GCE best-fit spectra (BB+NB components) reconstructed with the \SF\ fit in the range 10 -- 300 GeV (orange solid, baseline), testing uncertainties related to the input spectrum of the innermost gas ring (green, gas I); 
spectral freedom on the {\it Fermi} bubbles (royal blue, 25\% and navy, 50\%); spatial morphology of the {\it Fermi} bubbles
(teal, structured morphology). 
We also show the GCE Bulge spectrum when 
the new BB determination in Ref.~\cite{2020MNRAS.495.3350C} is adopted (red line, BB Coleman).
}
\end{figure*}

First, we consider the fit in the 0.3 -- 300 GeV. 
In figure~\ref{fig:SF_spectra_syst} (left), we show the best-fit spectra of the GCE Bulge component resulting from 
the \SF\ when assuming different GCE input spectra: (i) a power law with index $\alpha = -2.5$ (baseline of this work); 
a power law with exponential cutoff with parameters (ii) $\alpha = -1.46$ and $E_0 = 3.6$ GeV (as assumed in previous
analyses as a proxy for the cumulative MSP spectrum, and also adopted in \paperI) or (iii) $\alpha = -2$ and $E_0 = 50$ GeV.
The latter input spectrum is meant to provide significant input flux at high energies so to mimic the behavior 
of the pure power law.
From figure~\ref{fig:SF_spectra_syst} (left), we can conclude that the GCE spectral behavior is consistent regardless 
of the assumed input spectrum up to $E \sim 10$~GeV. At higher energies, the (ii) input spectrum is basically zero and cannot be 
properly modulated despite the full spectral freedom. 
Instead, input spectra with non-zero high-energy values, i.e.~like (i) and (iii), can be modulated also at higher energies.
The best-fit GCE spectra we find for (i) and (iii) are compatible above 10 GeV, showing that, as long as the input spectrum is
non-zero in the energy range of interest \SF\ successfully modulates it providing robust results against the choice of the
input spectral parametrization. 
At this point, we recall that the spectral and spatial terms of 
each model component are factorized in the fit to the data. We studied the same variations of the input spectrum also for the case of an NFW126 GCE morphology, finding very analogous energy behaviors of the best-fit GCE spectrum.
For all spectral choices, there is strong evidence for the GCE, and the Bulge is needed on top of the NFW126 component.

Secondly, we consider the high-energy interval 10 -- 300~GeV, and perform \SF\ fits therein.
As mentioned in section~\ref{sec:SF_results}, the evidence of the GCE at higher energies is still strong, and 
a Bulge component is still required on top of NFW126.
We here perform a series of runs to test different systematics.
First, we noticed that the innermost gas ring (gas I) resulted to be strongly 
suppressed with respect to the best-fit in the 0.3 -- 300 GeV \SF\ run.
We can explain that because of the fact that the higher energy data possess less constraining power on this 
component given the reduced statistics and the fact that the gas emission is the stronger at low energies, 
where the gamma-ray production from neutral pion decay is peaked. 
We setup a run in which the gas I component is therefore initialized to same best-fit value of the 0.3 -- 300 GeV \SF\ run.
The spectral behavior of the GCE component is stable against this variation, and the evidence for GCE and preference for Bulge at 
high energies are unaffected. We display the high-energy GCE Bulge best-fit spectrum in figure~\ref{fig:SF_spectra_syst} (right).

The modeling of the {\it Fermi} bubbles \cite{Herold:2019pei} was found to represent an important source of systematic 
uncertainty in the evidence of a high-energy tail of the GCE. In particular, Ref.~\cite{Fermi-LAT:2017opo}
found that whenever a low-latitude {\it Fermi} bubbles template was added to the fit, the GCE high-energy tail 
($E>10$ GeV) disappeared, see their figure 9, right panel.
We would like to stress here that the {\it Fermi} bubbles are an effective gamma-ray model component, defined in \SF\ through their spectrum. 
We constrain the spectrum 
to be compatible with the best-fit spectral behavior at high-latitude found in Ref.~\cite{Fermi-LAT:2014sfa}, which
is approximately $\propto E^{-1.9}$, and allow it to 
vary only mildly. On the other hand, the spatial morphology is completely reconstructed by \SF\ through modulation 
of the spatial coefficients which are fully free to vary, see details in Ref.~\cite{Storm:2017arh,Calore:2021jvg}.
Therefore, by construction, for all \SF\ runs the {\it Fermi} bubbles do possess a low-latitude component.

\begin{figure*}[t]
\includegraphics[width=0.45\textwidth]{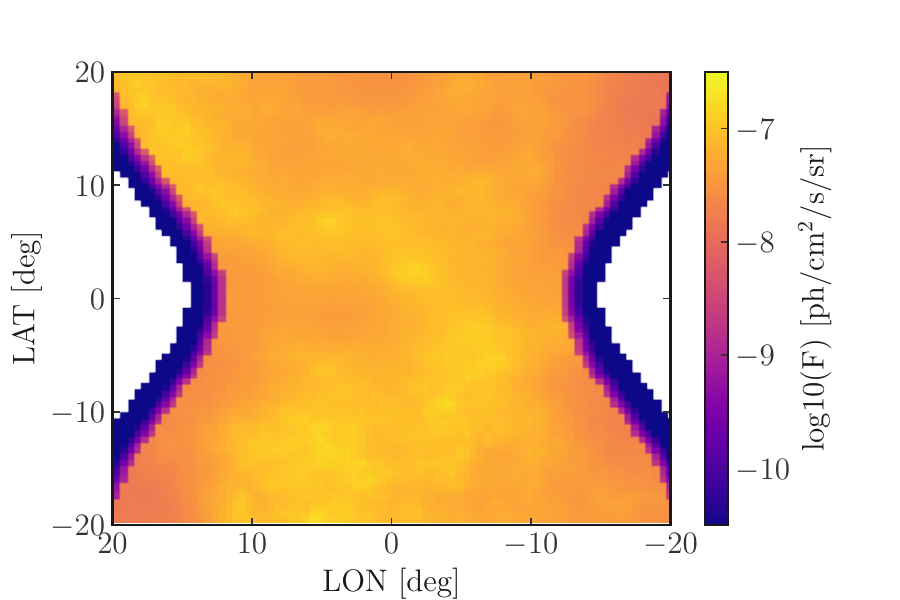}
\includegraphics[width=0.45\textwidth]{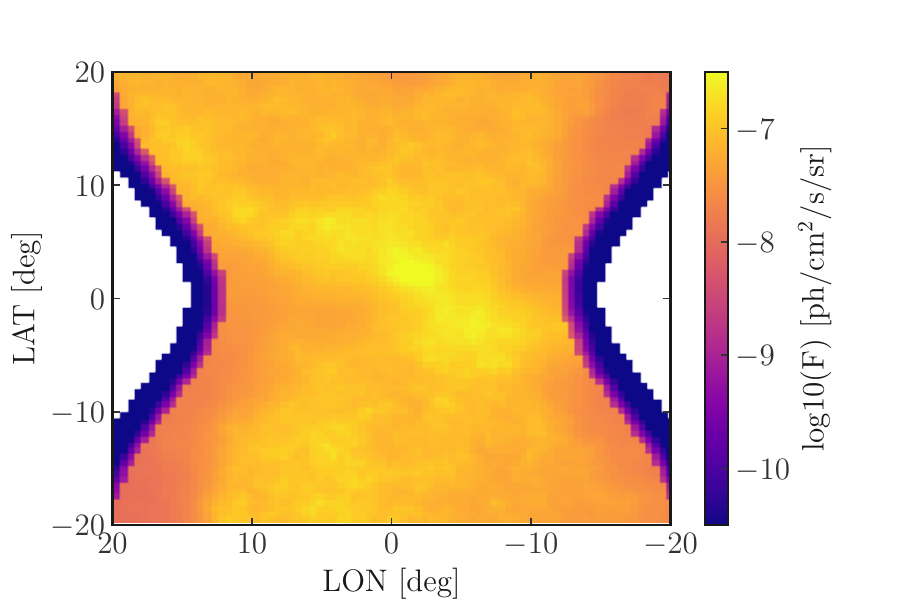}
\caption{\label{fig:FB_maps} \textit{The \SF-optimized templates} for the FB as integrated in the 10 -- 300~GeV energy bin and obtained from the fit in the full energy range (left) and at high energies (right).}
\end{figure*}

Nevertheless, given the possible strong degenerancies between {\it Fermi} bubbles and GCE, we test some possible systematics related to modeling of the {\it Fermi} bubbles.
In our framework, given the degenerancies at play, we cannot obtain convergent fit if both {\it Fermi} bubbles and GCE are left free spatially and spectrally, so 
that we need to constrain partly their model in order to get meaningful results. 
We therefore test the following setups: (a) We progressively increase the spectral freedom of the {\it Fermi} bubbles from the 1\% variation in the baseline setup
to 25\% and 50\%; (b) Instead of leaving the morphology of the {\it Fermi} bubbles free to readjust during the fit, we fix their spatial part to the 
best-fit structured template from Ref.~\cite{Macias:2019omb} and allow full spectral freedom. 
In figure~\ref{fig:SF_spectra_syst} (right), we show that the high-energy GCE spectrum is stable against variations 
of {\it Fermi} bubbles spectrum and morphology. 
The evidence for GCE is always present for these model variations, albeit slightly reduced (to 5.3 $\sigma$) when using a fixed, structured {\it Fermi} bubbles morphology,
which however provides overall a worse fit than our baseline setup. 
The \SF-optimized templates for the FB, corresponding to the fit performed in the full energy range (left) and for $E>10$~GeV (right) are shown for reference in figure~\ref{fig:FB_maps}.

Finally, we also test a more recent model of the stellar bulge~\cite{2020MNRAS.495.3350C},
which has been found to better fit the gamma-ray data in the full energy range~\cite{2024arXiv240102481Z,Song24prep}. 
The best-fit Bulge spectrum at high energy is fully consistent when different bulge templates are used.

\section{\Opp\ fits: systematics and fit ranges}
\label{app:opp_sys}

\begin{figure}[t]
\includegraphics[width=0.48\textwidth]{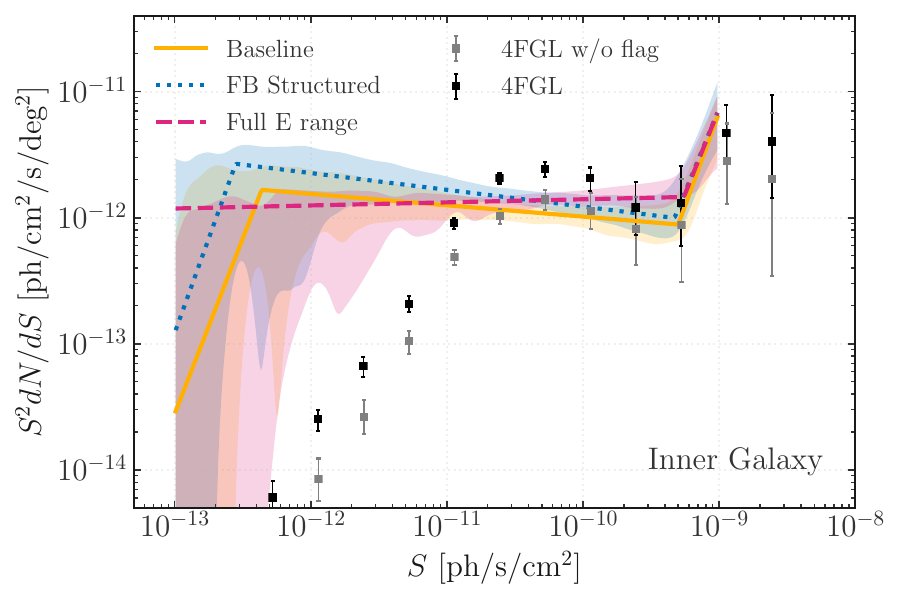}
\caption{\label{fig:op_spectra_syst} \textit{Systematic tests for the IG \dnds\ reconstruction using different \SF\  diffuse emission models. } 
The source count distribution of the IG (masking $b<|1|^\circ$) as obtained from the \Opp~analysis is measured varying various ingredients of the \SF\ Galactic diffuse emission derivation. The 'Full E range' (magenta dashed)  corresponds to the case in which all the diffuse templates are taken from the \SF\ fit in the 0.3--300~GeV interval. See the right panel of figure~\ref{fig:SF_spectra_syst} and text for more details. }
\end{figure}

To demonstrate that the most crucial  systematics on the \SF\ fit explored in the previous section have no impact on the \Opp\ results in the IG, we show in figure~\ref{fig:op_spectra_syst} the \dnds\ reconstructed for various cases. 
The results of the Baseline fit obtained in the 10--300~GeV  are found to be consistent with the ones obtained using the \SF\ diffuse models fitted to the full energy range (magenta dashed), as well as the ones using the structured FB template (blue dotted), no matter the latitude cut. 
Additionally, the $\ln(\mathcal{Z})$ for the fit using the \SF\ diffuse models obtained in the full 0.3--300~GeV range are respectively  -19118.29 and  -18119.29 for the latitude cut at $0.5^\circ$ and $1^\circ$, indicating that the \SF\ diffuse models optimized for the high energy fit better describe the gamma-ray sky fitted by the \Opp.

\begin{figure*}[t]
\includegraphics[width=0.48\textwidth]{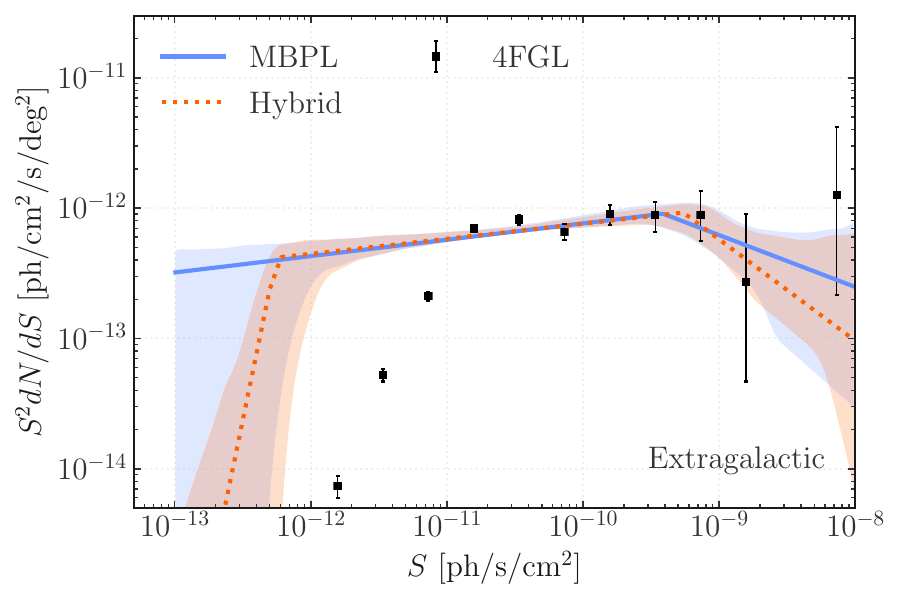}
\includegraphics[width=0.48\textwidth]{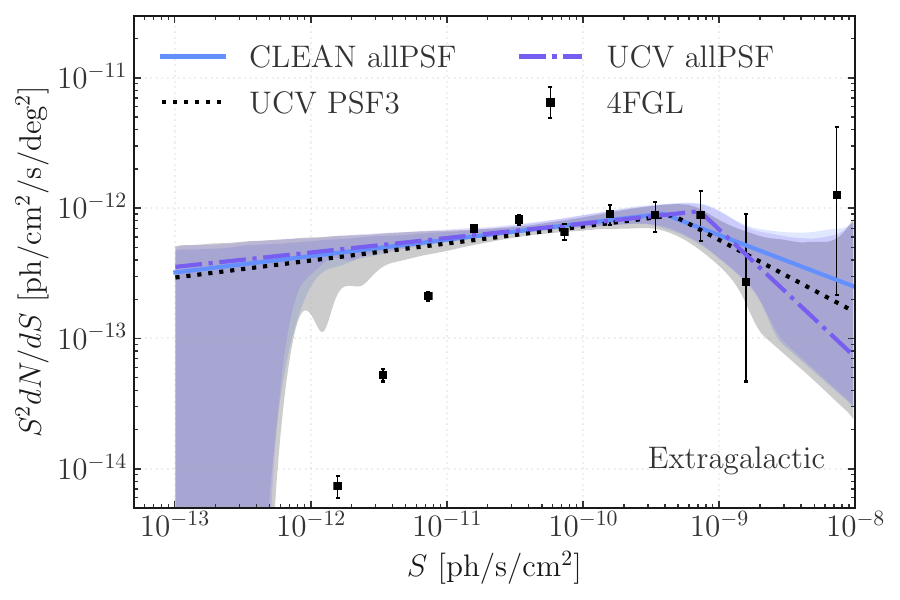}\\
\includegraphics[width=0.48\textwidth]{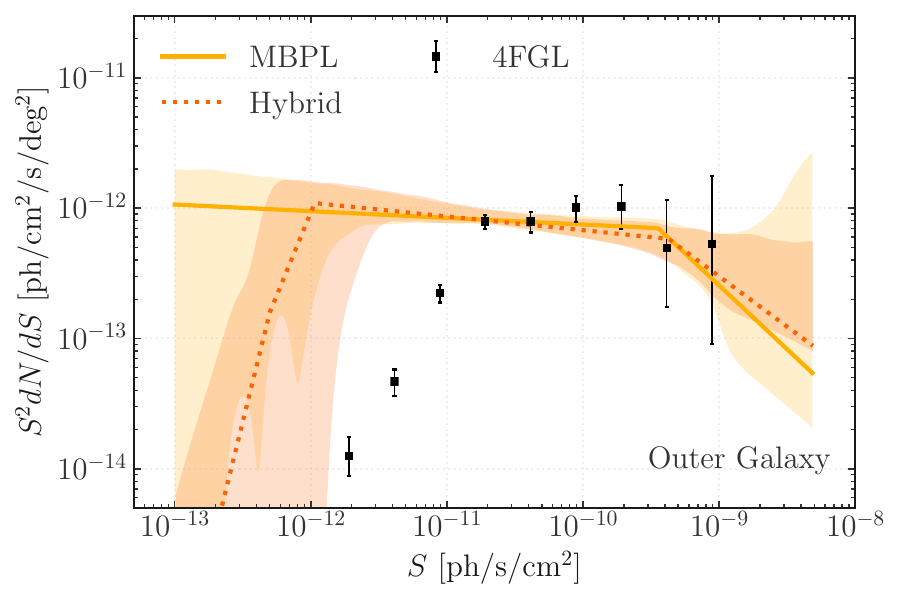}
\includegraphics[width=0.48\textwidth]{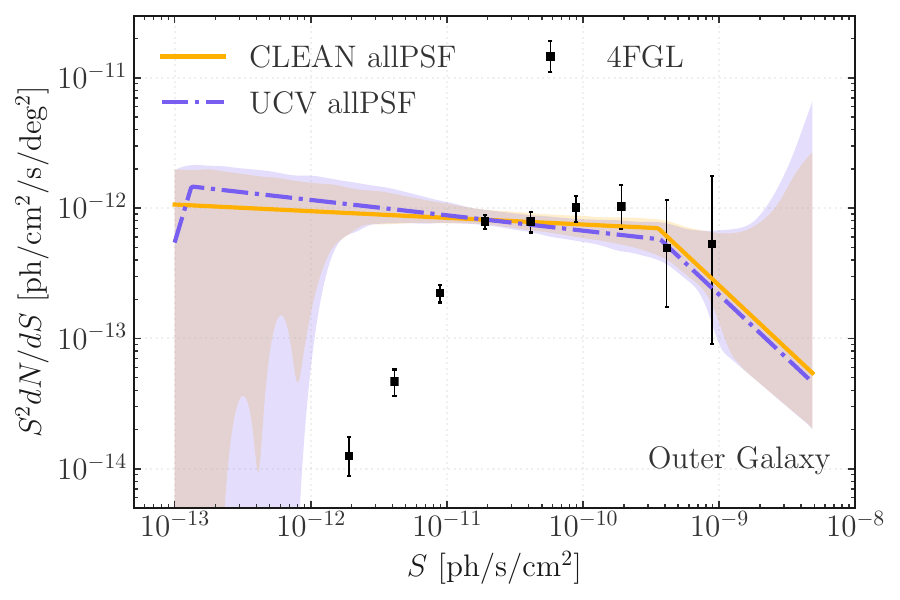}
\includegraphics[width=0.48\textwidth]{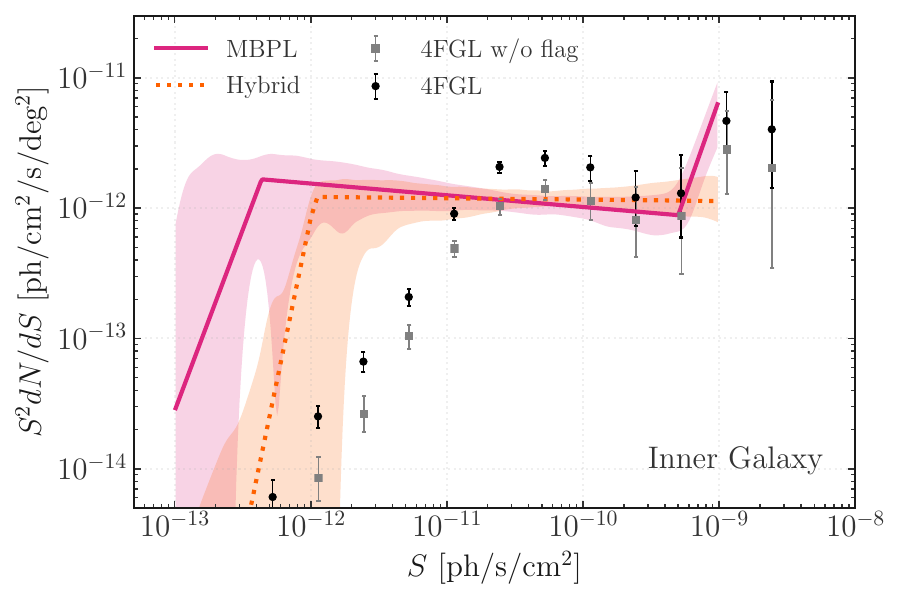}
\includegraphics[width=0.48\textwidth]{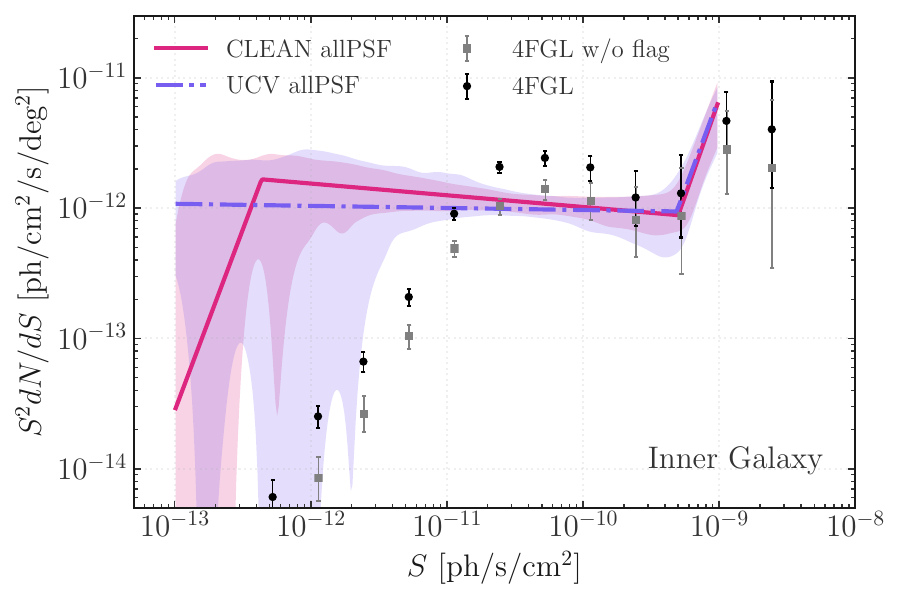}
\caption{\label{fig:EG_OG_IG_syst} \textit{  Systematic tests on \dnds~reconstruction.} 
Source count distribution of the extragalactic ROI (upper panels), of the outer Galaxy (center panels) and of the IG (masking $b<|1|^\circ$, and using the SF-BB setup) as obtained from the \Opp~analysis varying the \Opp~fit method (left) and the \textit{Fermi}--LAT event selection (right). Consistent results are obtained in all ROIs for all explored configurations. }
\end{figure*}

The panels in figure~\ref{fig:EG_OG_IG_syst} illustrate further systematics test performed on the reconstructed \dnds~ within the \Opp\ fits. 
The upper, middle and lower rows depict the results for the EG, OG and IG ROI, respectively. 
In each row, the left panel shows that the \Opp\ results presented in the main text are robust with respect of the modeling of faint point sources in the unresolved regime. Indeed, modeling the \dnds\ within the Hybrid method \cite{Zechlin:1}, thus adding nodes at fluxes $<10^{-12}$ \fluxunits leaves the \dnds~ results compatible with the MBPL fit down to the methods's sensitivity. 
The right panels of figure~\ref{fig:EG_OG_IG_syst} show instead how the \dnds\ results are robust against the choice of the \textit{Fermi}--LAT event selection. The measured \dnds\ is fully consistent within $1\sigma$ uncertainties among \texttt{CLEAN} and \texttt{UCV} events, and also restricting to the best quartile of PSF events (PSF3). The only observable effect is an increased  $1\sigma$ uncertainty band for event types including less photons, specifically in small ROIs such as the OG and the IG.

\begin{table}
\caption{\Opp~analysis of the IG: parameter priors. See text for details.  }
\centering
\begin{tabular}{ l @{\hspace{10px}}| l @{\hspace{10px}}| l @{\hspace{10px}}| c} \hline\hline
Method	& Parameter	& Prior& Range  \\ \hline
  & $A_{\rm gal}$	& flat & [0.5,1.5]\\ 
    	& $F_{\rm iso}$	& log-flat & [$4\cdot 10^{-9}$, $5 \cdot 10^{-8}$]\\ 
    		& $A_{\rm B/NFW126}$& flat & [$0.5$, $1.5$]\\ \hline
	
 MBPL   & $A_S$	& log-flat & [$10^{9}$, $5\cdot10^{13}$]\\ 
     	& $S_{\rm b1}$	& log-flat & [$3\cdot10^{-11}$, $10^{-9}$]\\ 
     	& $S_{\rm b2}$	& log-flat & [$1\cdot10^{-14}$, $3\cdot10^{-11}$] \\ 
     	& $n_{\rm 1}$	& flat & [-1, 3]\\ 
     	& $n_{\rm 2}$	& flat & [1.5, 2.5]\\ 
     	& $n_{\rm 3}$	& flat & [-2 , 2 ] \\  \hline
Hybrid  & $A_S$	& log-flat & [$10^{7}$, $10^{10}$]\\ 
     	& $S_{\rm b1}$	& log-flat & [$ 3\cdot 10^{-11}$, $10^{-9}$]\\ 
     	& $S_{\rm b2}$	& log-flat & [$10^{-12}$, $3\cdot 10^{-11}$]\\ 
     	& $n_{\rm 1}$	& flat  & [2.05 ,7]\\ 
     	& $n_{\rm 2}$	& flat  & [1.7,2.3]\\ 
     	& $n_{\rm 3}$	& flat  & [-2,2]\\  
     	& $A_{\rm nd1}$	& log-flat & [$10^{8}$, $10^{15}$]\\ 
     	& $S_{\rm nd1}$	& fixed & $10^{-13}$\\ 
     	& $n_{\rm f}$	& fixed & -10\\ \hline\hline
\end{tabular}
\label{tab:priors}
\end{table}

We summarize in Table~\ref{tab:priors} the parameter priors for the \Opp\ analysis of the IG. 
The table contains three blocks: in the first, parameters in common with all the fit setups are given, such as the normalization of the  diffuse emission templates and of the isotropic emission. 
The second block refers to the parameters for the \dnds~fit when using the MBPL approach, specifically the normalization of the \dnds\, the break positions and indexes, see Eq.~\eqref{eq:mbpl}.
The last block refers to the Hybrid approach, where the MBPL is extended with a node. The normalizations $A_S, \, A_{\rm nd1}$ are in units of  s~cm$^{2}$sr$^{-1}$, and the break positions $S_{\rm bn, snd1}$ are in units of \fluxunits. Finally, $F_{\rm iso}$ is given in units of cm$^{2}$s$^{-1}$sr$^{-1}$. All other parameters are dimensionless.

\end{document}